\documentclass[onecolumn,draftclsnofoot,10pt]{IEEEtran}
\usepackage{epstopdf}
\usepackage[draft]{hyperref} 
\usepackage{setspace}
\usepackage{amsmath}
\usepackage{amssymb}
\usepackage{amsthm}
\usepackage{amsfonts}
\usepackage{color}
\usepackage{graphicx}
\usepackage{subfigure}  
\usepackage[utf8]{inputenc}
\usepackage[T1]{fontenc}
\usepackage{cite}
\usepackage{mathtools}
\usepackage{stmaryrd}
\usepackage[mathscr]{euscript}
\usepackage{mathrsfs}

\usepackage{soul}

\usepackage{enumitem}

\usepackage{algorithm}
\usepackage{algpseudocode}

\begin{document}
	\author{  Satish~Mulleti, {\it Member, IEEE}, Haiyang Zhang, {\it Member, IEEE}, Yonina C. Eldar, {\it Fellow, IEEE}
		\thanks{\scriptsize All the authors are with the Faculty of Math and Computer Science, Weizmann Institute of Science, Israel. Email: satish.mulleti@gmail.com, haiyang.zhang@weizmann.ac.il, yonina.eldar@weizmann.ac.il}
		\thanks{\scriptsize This project has received funding from Israeli Council for Higher Education (CHE) via the Weizmann Data Science Research Center; European Union’s Horizon 2020 research and innovation program under grant No. 646804-ERC-COG-BNYQ; and the Israel Science Foundation under grant no. 0100101. }
		
	}
	\title{Learning to Sample: Data-Driven Sampling and Reconstruction of FRI Signals}
	\markboth{Submitted to the IEEE Transactions on Signal Processing}%
	{Shell \MakeLowercase{\textit{et al.}}: Bare Demo of IEEEtran.cls for Journals}
	\maketitle
	
	\begin{abstract}
		Finite-rate-of-innovation (FRI) signals are ubiquitous in applications such as radar, ultrasound, and time of flight imaging. Due to their finite degrees of freedom, FRI signals can be sampled at sub-Nyquist rates using appropriate sampling kernels and reconstructed using sparse-recovery algorithms. Typically, Fourier samples of the FRI signals are used for reconstruction. The reconstruction quality depends on the choice of Fourier samples and recovery method. In this paper, we consider jointly optimize the choice of Fourier samples and reconstruction parameters. Our framework is a combination of a greedy subsampling algorithm and a learning-based sparse recovery method. Unlike existing techniques, the proposed algorithm can flexibly handle changes in the sampling rate and does not suffer from differentiability issues during training. Importantly, exact knowledge of the FRI pulse is not required. Numerical results show that, for a given number of samples, the proposed joint design leads to lower reconstruction error for FRI signals compared to independent data-driven design methods for both noisy and clean samples. Our learning to sample approach can be readily applied to other sampling setups including compressed sensing problems. 
	\end{abstract}
	
	\IEEEpeerreviewmaketitle

	\section{Introduction}
	Sub-Nyquist approaches make use of signal structures beyond bandlimitedness \cite{eldar_2015sampling}. A popular structure is that of finite rate of innovation (FRI) which has been extensively studied due to its widespread applications in radar signal processing, ultrasound imaging, and time of flight imaging \cite{vetterli, bar_radar, bajwa_radar, mulleti_icip, eldar_sos}. A typical FRI signal consists of a stream of known pulses, where the information of the signal lies in the amplitude and time-delays of the pulses. For an FRI signal consisting of a stream of $L$ pulses there are $2L$ unknowns. It has been shown that a minimum of $2L$ Fourier samples of the FRI signal are sufficient to uniquely identify the unknowns in the absence of noise\cite{eldar_sos, fri_strang, mulleti_paley}. In practice, due to the limitation of the recovery algorithms or noisy measurements, more than $2L$ Fourier samples are needed to achieve good reconstruction accuracy.  
	
	The desired Fourier samples of the FRI signals can be computed by using a kernel-based sub-Nyquist sampling framework. In this framework, the FRI signal is passed through an appropriate sampling kernel prior to sampling. Then, the filtered signal is sampled at a rate proportional to the number of Fourier samples to be determined. The sampling rate is far below the effective bandwidth of the FRI signal, which results in sub-Nyquist sampling. The sampling kernel, which is a function of the desired Fourier samples, plays a crucial role in determining Fourier measurements of the FRI signal from its sub-Nyquist samples. The amplitude and time-delays of the FRI signal are then determined from the Fourier samples either by using high-resolution spectral estimation methods (HRSE) \cite[Ch. 4]{stoica} or by applying compressive sensing (CS) techniques \cite{eldar_cs_book}, such as orthogonal matching pursuit (OMP) \cite{omp}, iterative
	shrinkage/thresholding algorithms (ISTA) \cite{ista}, or its fast-counterpart, fast-ISTA (FISTA) \cite[Ch. 2]{eldar_optimization_book}, \cite{fista}.

	In the aforementioned standard sampling and recovery framework, for a fixed number of Fourier samples or sampling rate, reconstruction accuracy is a function of the choice of Fourier samples and the reconstruction algorithm. Hence it is desirable to optimize the subsampling of Fourier samples and parameters of the reconstruction algorithm to improve reconstruction accuracy for a fixed number of samples. For example, the spectrum of the FRI pulse plays a crucial role in the reconstruction algorithm and choice of Fourier coefficients, especially in the presence of noise. Moreover, in many applications, the sparsity pattern of the FRI signals is further structured, which in turn can be used to reduce the number of measurements below what is required by standard CS methods  \cite{baraniuk_model_cs}. To optimize the selection of the Fourier samples and reconstruction method, prior knowledge of the FRI pulse shape, the sparsity structure, and noise levels is required which may not always be available in practice. For example, in radar imaging, the FRI pulse denotes the transmit signal which may be distorted during transmission and not known at the receiver. 
	
	
	In the absence of such prior knowledge, data-driven techniques are well suited. In these approaches, a set of examples that are similar to the signal to be sampled are available. These examples implicitly carry information of the pulse shape, sparsity structure, and noise levels. Even when this information is known explicitly, there does not exist an optimal data-independent technique to solve the problem of Fourier sample selection. Data-driven reconstruction and subsampling methods learn hidden information by fine-tuning their parameters over the examples. In the context of subsampling existing data-driven approaches can be broadly classified into three categories: (1) data-driven recovery methods with a fixed subsampling pattern; (2) data-driven subsampling approaches with fixed recovery; and (3) joint subsampling and reconstruction methods. For example, deep-learning reconstruction techniques have been proposed to reduce the computational cost compared to standard CS algorithms and improve reconstruction accuracy in the presence of noise and sparsity structures \cite{distributed_deep_cs, lohit_deep_cs, lista}. Among these, learning-based ISTA or LISTA \cite{lista} is the most popular due to computational efficiency and its interpretable structure. In addition, due to its model-based structure, LISTA requires fewer examples to train compared to standard deep learning approaches \cite{unrolling_mag}. Recently, several alternative model-based networks have been proposed for sparse recovery \cite{amp_cs, pokala_firmnet, trainable_ista}. Unlike HRSE or standard CS-based methods, LISTA or other model-based networks do not require explicit knowledge of the FRI pulse shape. However, CS-based approaches, both data-driven and data-independent, assume that the time-delays of the FRI signals are on a grid. Leung at al \cite{leung_icassp20, leung_eusipco21} proposed a data driven reconstruction technique for FRI signals where the time-delays are not necessarily on a grid. They show a higher reconstruction accuracy compared to standard HRSE techniques at low signal-to-noise ratio (SNR). In these techniques, the reconstruction parameters are optimized for a fixed sampling pattern and sampling kernel, and knowledge of the FRI pulse is required.

	Data-driven subsampling has been considered in \cite{Adcock2015,lustig_mri,ravishanker, oedipus, mulleti_radar, baldassarre,  gozcu}. These methods result in lower reconstruction error compared to  non-adaptive, random subsampling \cite{candes_spmag, candes_uncertainity}. Adcock et al. \cite{Adcock2015} considered a deterministic approach for subsampling Fourier measurements with an assumption that the signals are sparse in the wavelet domain. Lustig et al. \cite{lustig_mri} used a random variable density subsampling scheme for magnetic resonance imaging (MRI) where low frequencies are favoured. In \cite{oedipus, mulleti_radar, baldassarre,  gozcu}, greedy algorithms are used to design subsampling patterns. In \cite{oedipus, mulleti_radar} a Cr\'amer-Rao
	lower bound (CRLB)-based cost function, which is independent of any recovery method, is used to determine the subsampling pattern. In \cite{baldassarre,  gozcu} a means-squared error cost function is used which is computed assuming a fixed reconstruction method. In a direct application of these methods to the Fourier subsampling problem for FRI signals may not be feasible due to difference in the measurement model. Even if a modification of these methods suits the problem at hand, it requires knowledge of the FRI pulse shape.

	Compared to independent subsampling and reconstruction optimization methods, joint design results in lower reconstruction error for a given number of measurements as shown in the context of MRI \cite{yi_mri,  jmodl}, and ultrasound imaging \cite{huijben2019learning}. These methods consider both subsampling and reconstruction as parametric functions and optimize the parameters over a given data set such that reconstruction error is minimized. These joint design methods can be applied to select Fourier samples and reconstruction parameters for FRI signals. However, these techniques have several drawbacks. First, the approaches require exact knowledge of the measurement matrix which is a function of the pulse shape in the FRI context. Second, in the joint design works the cost function is not differentiable with respect to the sampling pattern, so that gradient-based optimization methods are not applicable. In \cite{jmodl}, the sampling patterns are constrained to be a continuous function of a known sampling template to address the differentiability issue. However, the selection of a sampling template is an additional challenge. In \cite{yi_mri,huijben2019learning}, categorical random variables or probabilistic policies are used to reparameterize the discrete sampling locations to a continuous random variable. This relaxation results in additional parameters to be learned and increases the computational cost. In addition, it is difficult to understand the effect of relaxation on the performance of joint optimization. Another key issue of these approaches is their lack of flexibility to changes in the subsampling rate. The networks in \cite{huijben2019learning, jmodl} requires retraining from scratch for changes in the sampling rate.

	In this paper, we propose a data-driven, model-based, deep learning approach to jointly design the sampling and reconstruction of FRI signals. Our method does not suffer from  differentiability issue and can be used to design a sampling kernel for FRI signals. Two main ingredients of our technique are a greedy algorithm that is used for subsampling the Fourier measurements and a LISTA network that is applied for recovery. At any given iteration of the greedy algorithm, a new sample is selected (or removed) from the remaining (or existing) samples such that it results in the lowest reconstruction error. The reconstruction error is computed from the ground truth and estimation of the sparse vector by a trained LISTA network. In our approach, at each iteration of the greedy step, reconstruction parameters are optimized for sampling patterns at that step. In this way, at the end of the final iteration, an optimal choice of Fourier samples, along with a matched reconstruction method is obtained. 
	
	Unlike existing joint approaches where the networks are optimized over two sets of parameters, one for sampling and one for recovery, in our solution, we only optimize over recovery parameters while the greedy steps aid subsampling. Hence, our approach does not have any differentiability issues. Furthermore, due to the use of the sequential greedy algorithm, our approach can start from the current sampling pattern and does not require retraining when the sampling rate changes. The selected Fourier samples are used to design a sampling kernel for FRI signals. Specifically, we use a sum-of-sincs (SoS) sampling kernel where the location of sincs in the frequency-domain are the same as the locations of the selected Fourier samples \cite{eldar_sos, mulleti_paley}. Application of the proposed joint subsampling and reconstruction approach is not limited to FRI signals but can be extended to other subsampling problems such as MRI, compressive sensing, and more.

	
	
We compare the proposed algorithm with existing methods in which sampling and recovery are independent of each other. In particular, given the greedy nature of our algorithm, we compare our results with greedy methods as in \cite{oedipus, mulleti_radar, baldassarre,  gozcu}. These methods require knowledge of the pulse shape. We show that our approach results in a lower error for both noisy and noise free cases. In particular, we show that our joint learning-based approach can estimate $L$ sparse vectors from less than $2L$ Fourier measurements up to $-20$ dB normalized mean-squared error (NMSE) in the absence of noise. In other words, one can consider sampling below the sub-Nyquist rate of FRI signals by using data-driven joint designs.
In addition, we show that the learning-based approach can learn sparsity structures in the signal by optimizing the sampling and reconstruction for a given set of examples.

	The paper is organized as follows. The problem of jointly designing a sampling kernel and reconstruction technique for FRI signals is discussed in Section~\ref{sec:problem_formulation}. In Section~\ref{sec:jsr}, we present the proposed joint-sampling and reconstruction algorithm. Numerical results are presented in Section~\ref{sec:numerical_results} followed conclusions in Section~\ref{sec:dis_conclusion}.

	\section{Problem Formulation}
	\label{sec:problem_formulation}
	\subsection{Signal Model}
	\label{sec:signal_model}
	Consider an FRI signal $f(t)$ that consists of a sum of $L$ amplitude-scaled and time-shifted copies of a pulse $h(t)$:
	\begin{align}
	f(t) = \sum_{\ell=1}^L a_\ell h(t-t_\ell).
	\label{eq:fri_model}
	\end{align}
	In the FRI framework, $h(t)$ is assumed to be known, which results in a finite degrees of freedom of the FRI signal $f(t)$ in terms of the amplitudes $\{a_\ell\}_{\ell=1}^L$ and delays $\{t_\ell\}_{\ell=1}^L$. FRI signal sampling and reconstruction deals with estimating the parameters $\{a_\ell, t_\ell\}_{\ell=1}^L$ from the fewest possible samples measured at a sub-Nyquist rate.

	In radar, ultrasound, and other time-of-flight-based applications, the pulse $h(t)$ denotes the transmit pulse. Assuming there are $L$ point targets in the medium, the transmit pulse gets reflected from the targets and a linear combination of $L$ reflected pulses is received. The received signal is modeled as the FRI signal in \eqref{eq:fri_model} where the amplitudes $\{a_\ell\}_{\ell=1}^L$ and $\{t_\ell\}_{\ell=1}^L$ denote area or size of the targets and location of the targets, respectively. 
	
	We make the following assumptions on the signal model. We assume that the number of targets, $L$, is known and, $t_\ell \in (0, t_{\max}]$, where $t_{\max}$ is the maximum time delay of the targets. In practice, the pulse $h(t)$ has compact support and we assume that $h(t) = 0$ for $t \not \in [0, T_h]$. The support assumption of $h(t)$ together with the assumption that $t_\ell \in (0, t_{\max}]$ results in compactly supported $f(t)$ over the interval $[0, T_h+t_{\max}]$.

	\begin{figure}[!t]
		\centering
		\includegraphics[width=3.5in]{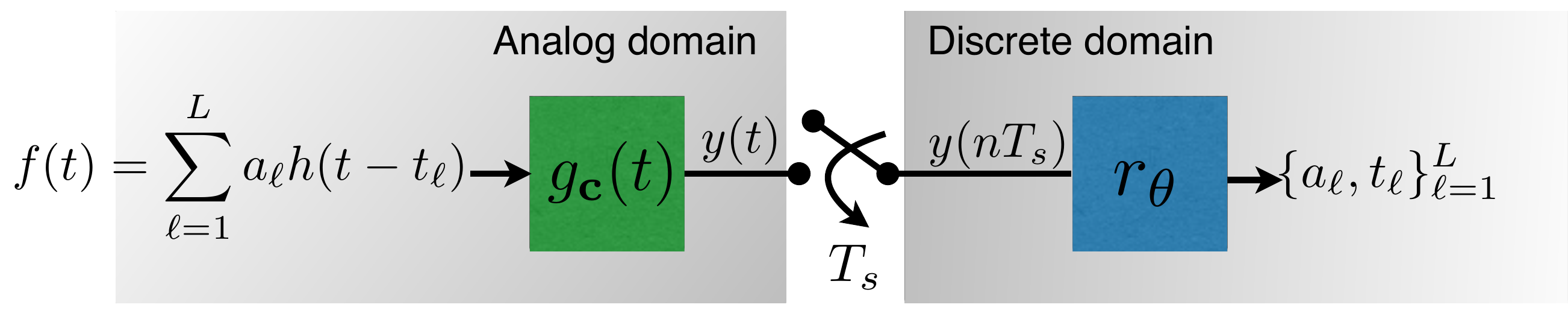}
		\caption{A schematic of FRI signal sampling and reconstruction: An FRI signal $f(t)$ is first passed through a filter $g_{\mathbf{c}}(t)$  where $\mathbf{c}$ is a parameter of the filter. The signal is uniformly sampled at a sub-Nyquist rate and the signal parameters are estimated by applying a reconstruction method $r_{\boldsymbol{\theta}}$.}
		\label{fig:fri_sampling}
	\end{figure}
	A schematic of a typical FRI signal sampling and reconstruction framework is shown in Fig.~\ref{fig:fri_sampling}. The sampling process consists of a sampling kernel $g_{\mathbf{c}}(t)$ followed by a sampler operating at a rate $\frac{1}{T_s}$ samples per second. The output of the sampler is a sequence of samples $y(nT_s)$ from which the FRI parameters $\{a_\ell, t_\ell\}_{\ell=1}^L$ are computed by the reconstruction method $r_{\boldsymbol{\theta}}$. The sampling kernel and reconstruction method are parameterized by $\mathbf{c}$ and $\boldsymbol{\theta}$, respectively. Our goal is to jointly design a sampling kernel and reconstruction method such that the FRI signal parameters are determined uniquely from sub-Nyquist samples. Specifically, for a given set of FRI signals of the form \eqref{eq:fri_model}, we seek to jointly optimize $\mathbf{c}$ and $\boldsymbol{\theta}$ to minimize the reconstruction error while operating at sub-Nyquist rates. Details of the problem formulation together with the sampling and reconstruction methods are discussed next.  
	
	\subsection{Frequency-Domain Reconstruction}
	\label{sec:sampling_recons}
	
	The reconstruction is typically carried out in the frequency domain as it facilitates application of either HRSE or CS-based methods. To elaborate, consider uniform Fourier-domain samples of $f(t)$ as
	\begin{align}
	F(k\omega_0) = H(k\omega_0) \sum_{\ell=1}^L a_\ell e^{\mathrm{j}k\omega_0 t_\ell}, \quad k \in \mathcal{N},
	\label{eq:fri_fourier}
	\end{align}  
	where $\omega_0 \in \mathbb{R}$ is the sampling interval and $\mathcal{N}$ is a set of integers. The sampling locations are chosen such that $\{H(k\omega_0)\}_{k \in \mathcal{N}}$ do not vanish. Since $\{H(k\omega_0)\}_{k\in \mathcal{N}}$ are computed a priori from the known pulse $h(t)$, from the Fourier samples the following sequence of samples are derived:
	\begin{align}
	S(k\omega_0) = \frac{F(k\omega_0)}{H(k\omega_0)} =\sum_{\ell=1}^L a_\ell e^{\mathrm{j}k\omega_0 t_\ell}, \quad k \in \mathcal{N}.
	\label{eq:SoWe}
	\end{align}  
	The spectral samples $S(k\omega_0)$ are a function of the unknown FRI parameters $\{a_\ell, t_\ell\}_{\ell=1}^L$.  
	
	To estimate the FRI parameters through HRSE methods, such as annihilating filter \cite{prony} or matrix-pencil algorithms \cite{sarkar_mp}, the set $\mathcal{N}$ should consist of consecutive integers with cardinality $|\mathcal{N}|\geq 2L$. In other words, there should be $2L$ spectral samples to determine $2L$ unknowns. In addition, for uniquely identifying the time-delays it is necessary that the elements in the $\{e^{\mathrm{j}\omega_0 t_\ell} \}_{\ell=1}^L$ are distinct. A straightforward way to ensure this requirement is to set $\omega_0 = \frac{2\pi}{t_{\max}}$.
	
	Compared to HRSE methods, CS-based algorithms, such as orthogonal matching pursuit or iterative thresholding techniques,  are shown to be robust in the presence of noise. This motivates us to consider CS-based algorithms for reconstruction in this paper. To apply these methods, the time-delays are assumed to be on a grid. To elaborate, let $\Delta$ be a grid size and $N = \frac{t_{\max}}{\Delta}$. Then for each $t_\ell$ there exists an integer $n_\ell \in \mathcal{N} = \{1, 2, \cdots, N\}$ such that $t_\ell = n_\ell \Delta $. By substituting the grid assumption in \eqref{eq:fri_fourier} together with $\omega_0 = \frac{2\pi}{t_{\max}}$, we have
	\begin{align}
	F(k\omega_0)  = H(k\omega_0)\sum_{n=1}^N x_n e^{\mathrm{j}2\pi kn/N}, \quad k \in \mathcal{N},
	\label{eq:SoWe_grid}
	\end{align} 
	where $\{x_{n_\ell} = a_\ell\}_{\ell=1}^L$ and $x_{n} = 0$ for $n \not\in \{n_\ell\}_{\ell=1}^L$. Within the assumed settings, the Fourier samples in \eqref{eq:fri_fourier} can be written in matrix notation as
	\begin{align}
	\mathbf{f} = \text{diag}(\mathbf{h}) \mathbf{A} \mathbf{x},
	\label{eq:fri_fourier1}
	\end{align}  
	where $\mathbf{f} = [F(\omega_0), F(2\omega_0), \cdots, F(N\omega_0)]^{\mathrm{T}}$, $\mathbf{h} = [H(\omega_0), H(2\omega_0), \cdots, H(N\omega_0)]^{\mathrm{T}}$, and $\mathbf{A}$ is an $N$-th order discrete Fourier transform (DFT) matrix. The vector $\mathbf{x} = [x_1, \cdots, x_N]^{\mathrm{T}}$ is $L$-sparse where its support carries the information of time-delays and the values of its non-zero entries are equal to the amplitudes of the FRI signal. To estimate $\mathbf{x}$, a minimum of $2L$ samples of $\mathbf{f}$ are required in the absence of noise and more than $2L$ in the presence of noise. 
	
	\subsection{Kernel-Based Sampling}	
	For both on-grid or gridless premises, the frequency-domain-based reconstruction requires the computation of Fourier samples $\{F(k\omega_0)\}_{k \in \mathcal{N}}$. This is achieved through carefully designing the sampling kernel and the sampling rate. In the context of FRI sampling, different sampling kernels are proposed, such as SoS kernels, generalized kernels, Gaussian, ideal lowpass filters, exponential and polynomial generating kernels, and more. Among these, the SoS kernel, the generalized kernels, and the polynomial generating kernels are suitable for the present discussion. These kernels can be designed such that the filtered signal $y(t)$ is a linear combination of Fourier samples of $f(t)$ as required in the Fourier-domain recovery method. Moving forward, we consider an SoS kernel due to its simplicity in design and parametric nature. 
	
	An SoS kernels has compact support and its impulse response is given by
	\begin{align}
	g_{\mathbf{c}}(t) = \sum_{k\in \mathcal{N}} c_k e^{\mathrm{j}k \omega_0 t}, \quad t \in [0, T_g], 
	\label{eq:sos} 
	\end{align} 
	where $T_g> T_h+t_{\max}$. The coefficients $\{c_k\}_{k\in \mathcal{N}}$ are design parameters that can be designed to select specific Fourier coefficients as explained next. 
	
	It can be shown that the filtered signal $y(t)$ by using the SoS filter is given as
	\begin{align}
	y(t) = \sum_{k \in \mathcal{N}} c_k F(k\omega_0) e^{\mathrm{j}k\omega_0 t}, \quad t \in [T_h+t_{\max}, T_g].
	\label{y_sos}
	\end{align}
	Upon uniformly sampling $y(t)$ in the interval $[T_h+t_{\max}, T_g]$ we have the following samples
	\begin{align}
	y(nT_s) = \sum_{k \in \mathcal{N}} c_k F(k\omega_0) e^{\mathrm{j}k\omega_0 nT_s}.
	\label{eq:ynTs}
	\end{align}
	To uniquely determine scaled Fourier samples $\{F(k\omega_0)\}_{k\in \mathcal{N}}$  from the time-samples in \eqref{eq:ynTs}, there should be atleast $|\mathcal{N}|$ time-samples and the set of linear equations in \eqref{eq:ynTs} should be invertible. By choosing $T_g \geq T_h+t_{\max}+|\mathcal{N}|T_s$, it can be ensured that there are at least $|\mathcal{N}|$ time-samples. The linear equations are invertible provided that $T_s = \frac{t_{\max}}{|\mathcal{N}|}$. Note that the sampling rate is proportional to the number of Fourier samples.
	
	In practice, $N \gg L$ to ensure that the on-grid assumption is close to any off-grid scenario. The condition $N \gg L$ results in a large number of time and Fourier samples beyond the minimal $2L$ samples. As the sampling rate is proportional to the number of Fourier samples, $N \gg L$  leads to over sampling.
	
	To ensure that the desired number\footnote{$2L$ in the absence of noise and more than $2L$ in the presence of noise.} of Fourier samples are selected while sampling at the lowest possible rate Fourier-domain subsampling is required. Specifically, let $\mathcal{K} \subset \mathcal{N}$ where $2L\leq |\mathcal{K}|< |\mathcal{N}|$ denotes the index of Fourier samples in $\mathbf{\bar{f}}$ that are considered for reconstruction. One of the simplest ways to consider these Fourier samples is to set $c_k = 1, k \in \mathcal{K}$ and the remaining coefficients to be zero. The non-zero values of $c_k$ can be designed to take any value instead of $c_k = 1$. However, hardware implementation of the filter for arbitrary choice of $c_k$s may give rise to calibration issues. Hence for the reminder of the discussion we consider binary valued $c_k$s.

	With subsampling, \eqref{eq:ynTs} becomes
	\begin{align}
	y(nT_s) = \sum_{k \in \mathcal{K}} c_k F(k\omega_0) e^{\mathrm{j}k\omega_0 nT_s}, \quad n \in \mathcal{N},
	\label{eq:ynTs2}
	\end{align}
	where the summation is over $\mathcal{K}$ instead of $\mathcal{N}$. For reconstruction, $\{F(k\omega_0)\}_{k \in \mathcal{K}}$ needs to be determined from the time samples in \eqref{eq:ynTs2}. In the set of linear-equations in \eqref{eq:ynTs2}, the matrix relating the Fourier and time samples is a Vandermonde matrix with its $(nk)$-th component being $e^{\mathrm{j}k\omega_0 n T_s}$ where $n \in \mathcal{N}$ and $k \in \mathcal{K}$. The Vandermonde matrix is left-invertible provided that its seeds $\{e^{k\omega_0 T_s}\}_{k \in \mathcal{K}}$ are distinct. Since $\mathcal{K} \subset \mathcal{N}$, the choice $T_s = \frac{t_{\max}}{|\mathcal{N}|}$ ensures uniqueness of recovery. However, this results in higher sampling rate. 
	
	To ensure that the sampling rate is proportional to the number of Fourier samples $|\mathcal{K}|$, we set the sampling interval as 
	\begin{align}
	T_s = \frac{t_{\max}}{|\mathcal{K}|+\epsilon},
	\label{eq:Ts}	
	\end{align}
	where $1> \epsilon>0$. The condition $\epsilon = 0$ should give a minimum desirable sampling rate, however, this may result in repetition in the set $\{e^{\mathrm{j}k\omega_0 T_s}\}_{k \in \mathcal{K}}$.
	For example, if $k_0, k_0+|\mathcal{K}| \in \mathcal{K}$ for some integer $k_0$, then the seeds $e^{\mathrm{j}k_0\omega_0 T_s}$ and $e^{\mathrm{j}(k_0+|\mathcal{K}| )\omega_0 T_s}$ are indistinguishable for $\epsilon = 0$.	The upper bound, $\epsilon<1$, ensures that the sampling rate is close to the desirable minimum rate while ensuring distinct seeds.

	\subsection{Problem of Joint Sampling and Recovery}
	The Fourier measurements computed through SoS filtering can be written in the following compact form:
	\begin{align}
	\mathbf{\bar{f}} = \text{diag}(\mathbf{c})\mathbf{f}= \text{diag}(\mathbf{c}) \text{diag}(\mathbf{h}) \mathbf{A} \mathbf{x},
	\label{eq:fri_fourier2}
	\end{align}  
	where $\mathbf{c}$ consists of the coefficients of the SoS kernel. Note that $\mathbf{c} \in \{0, 1\}^N$ and the support of $\mathbf{c}$ is $\mathcal{K}$, that is, $\text{supp}(\mathbf{c}) = \mathcal{K}$. From $\mathbf{\bar{f}}  = \text{diag}(\mathbf{c})\text{diag}(\mathbf{h})\mathbf{Ax}$, the sparse signal $\mathbf{x}$ is estimated as $\hat{\mathbf{x}} = r_{\boldsymbol{\theta}}\left(\text{diag}(\mathbf{c})\text{diag}(\mathbf{h})\mathbf{Ax}\right)$ where we assume that $\boldsymbol{\theta} \in \mathbb{C}^M$. The estimate $\hat{\mathbf{x}}$ is a function of the reconstruction parameters $\boldsymbol{\theta}$ and sampling kernel parameter $\mathbf{c}$ or alternatively, the choice of the Fourier samples set $\mathcal{K}$. It is desirable to choose $\mathbf{c}$ and $\boldsymbol{\theta}$ to minimize the error in estimation of $\mathbf{x}$. Instead of optimizing the sampling kernel and reconstruction for all possible choices of $\mathbf{x}$ and $\mathbf{h}$, taking inspiration from deep learning, we consider a data-driven approach to optimize $\mathbf{c}$ and $\boldsymbol{\theta}$. 
	
	For a given application, let a set of representative examples with measurement noise be available. Specifically, consider a data set 
	\begin{align}
	\mathcal{D} = \{\mathbf{f}_q =   \text{diag}(\mathbf{h}) \mathbf{A} \mathbf{x}_q, \mathbf{x}_q + \boldsymbol{ \eta}_q \}_{q=1}^Q
	\label{eq:data}
	\end{align}
	for fixed $\mathbf{h}$. The vector $ \boldsymbol{ \eta}_q$ denotes measurement noise in the $q$-th example. Our goal is to choose $\mathbf{c}$ and $\boldsymbol{\theta}$ to minimize the reconstruction error for the sparse vectors from $\mathcal{D}$. Specifically, we consider the  following optimization problem
	\begin{align}
	\underset{\substack{ \mathbf{c} \in  \{0, 1\}^{N}  \\ \boldsymbol{\theta} \in \mathbb{C}^M }}{\min}  & \quad  \frac{1}{Q} \sum_{q=1}^Q \{\mathrm{C}(\mathbf{x}_q, r_{\boldsymbol{\theta}} \left(\text{diag}(\mathbf{c})\mathbf{f}_q\right)\} \quad \text{s. t.}    \quad  \|\mathbf{c}\|_{0} = K, 
	\label{eq:opt_joint2}
	\end{align} 
	where $K$ is number of samples to be selected for reconstruction and the cost function $\mathrm{C}: \mathbb{C}^N \times \mathbb{C}^N \rightarrow [0, \infty)$  is a measure of reconstruction error. In these representative examples, in addition to being $L$-sparse, the examples could have structured sparsity. For example, the non-zero values could be bunched together or can occur in a specific pattern. These structures are not known a priori. Any solution to the optimization problem of the representative examples should be able to learn these structures and optimize $\mathbf{c}$ and $\mathcal{K}$ accordingly. From the solution to the optimization problem, a sampling kernel and recovery technique are designed that can be used to sample FRI signals whose sparsity structure and pulse shape are the same as the representative data set. Since the sampling kernel operates in an analog domain and the reconstruction is discrete, the above design principle is hybrid in nature. 	
	In the next section, we propose solutions to the problem discussed here.

	\section{Joint Sampling and Reconstruction Approach}	
	\label{sec:jsr}	
	In this section, we propose algorithms to solve the optimization problem in \eqref{eq:opt_joint2}. The optimization problem concerning $\mathbf{c}$ is combinatorial. To address this issue, we follow a greedy approach. Specifically, the sampling problem is solved by a non-parametric, iterative, greedy method. For reconstruction, we apply the LISTA \cite{lista}. To solve the sampling and reconstruction problem jointly, we combine the greedy algorithm and LISTA. Before discussing the proposed joint learning algorithm, we first discuss the standard greedy algorithm and its variants suitable for our objective and the LISTA algorithm. 

	\subsection{Greedy Algorithm for Sampling}
	Let us reconsider the optimization problem \eqref{eq:opt_joint2} where the objective is to select only an optimal sampling pattern $\mathbf{c}$ for a fixed recovery. To address the combinatorial nature of the problem convex relaxation approaches \cite{antennaSelectionViaCO, chepuri2015} or greedy methods have been suggested \cite{nemhauser,antennaSelectionKnapsack,mulleti_radar} in the context of array selection. Among these two methods, the greedy one is computationally efficient. In the greedy algorithm, the sampling pattern is computed in $K$-steps, and the performance of this method is close to optimal if the cost function is submodular \cite{fujishige2005submodular}.

	In \eqref{eq:opt_joint2}, for a fixed reconstruction method $r_{\boldsymbol{\theta}}$, the cost is a function of the set $\mathcal{K}$. Let 
	\begin{align}
	\mathrm{S}(\mathcal{K}) = \frac{1}{Q} \sum_{q=1}^Q \mathrm{C}(\mathbf{x}_q, r_{\boldsymbol{\theta}} \left(\text{diag}(\mathbf{c})\mathbf{f}_q\right).
	\label{eq:Scost}
	\end{align}
	The set function $\mathrm{S}(\mathcal{K})$ is submodular if it satisfies the property of decreasing marginals. Mathematically,  $\forall \mathcal{K}_1 \subseteq \mathcal{K}_2 \subseteq \mathcal{N}$ we have that $\mathrm{S}(\mathcal{K}_1\cup \{i\}) \geq \mathrm{S}(\mathcal{K}_2\cup \{i\}), \forall i \in \mathcal{N}\backslash \mathcal{K}_2$. The cost is said to be monotonic if $\forall \mathcal{K}_1 \subseteq \mathcal{K}_2 \subseteq \mathcal{N}$ we have that $\mathrm{S}(\mathcal{K}_1) \leq \mathrm{S}(\mathcal{K}_2)$ \cite{fujishige2005submodular}. The sampling problem,
	\begin{equation}
	\underset{\mathcal{K} \subset \mathcal{N} }{\min}  \quad \mathrm{S}(\mathcal{K}) \quad \text{s. t.}  \quad |\mathcal{K}| = K,
	\label{eq:jsr_opt3}
	\end{equation}  
	can be solved by a greedy approach in $K$-steps provided that the cost is submodular and monotone. The performance of the greedy approach reaches within $\left(1-\frac{1}{e} \right)$ of the optimal solution \cite{nemhauser}.
	
	Typically, a greedy algorithm starts with zero samples. At each iteration, a new sample is selected from the remaining samples that together with the selected one minimize the cost. The steps of the greedy approach are summarized in Algorithm~\ref{algo:greedy}.
	\begin{algorithm}[!t]
		\caption{Greedy Algorithm}
		\label{algo:greedy}
		\begin{algorithmic}
			\State 
			\State {\bf Initialize:} $\mathcal{K} =\emptyset$
			\For{$k=1$ to $K$ } 
			\State[S1]   $i ^{*}  = \underset{i \in \mathcal{N}\backslash \mathcal{K}}{\min} \quad \mathcal{S}(\mathcal{K} \cup \{i\})$  \\
			\State[S2] $\mathcal{K} = \mathcal{K} \cup \{i^*\}$
			\EndFor
		\end{algorithmic}
	\end{algorithm}
	
	Among several choices of submodular cost functions, cost as the log-determinant of the CRLB (log-det CRLB) is shown empirically to have small error while estimating sparse vectors from subsampled Fourier measurements \cite{mulleti_radar}. In the present setup, assuming that $\{\mathbf{x}_q\}_{q=1}^Q$ are deterministic, the CRLB in estimation of $\mathbf{x}_q$ from $\mathbf{y}_q  = \mathbf{B}\mathbf{x}_q+\boldsymbol{\eta}_q$ where $\mathbf{B} = \text{diag}(\mathbf{c}) \text{diag}(\mathbf{h}) \mathbf{A}$. Here CRLB or log-det CRLB is a function of the sampling pattern $\mathbf{c}$ or $\mathcal{K}$ as well as the pulse $\mathbf{h}$. Let $\text{CRLB}_q(\mathbf{x}_q, \mathbf{h}, \mathbf{c})$ denote the CRLB matrix in the estimation of $\mathbf{x}_q$, then the cost function in \eqref{eq:Scost} is given as 
	\begin{align}
	    \mathcal{S}(\mathcal{K}) = \frac{1}{Q} \sum_{q=1}^Q \text{log-det}\,\text{CRLB}_q(\mathbf{x}_q, \mathbf{h}, \mathbf{c}). 
	    \label{eq:crlb_cost}
	\end{align}
A similar CRLB-based cost function has been considered in \cite{oedipus} for subsampling using a greedy technique.

While the CRLB-based greedy algorithms work, the selected sampling pattern is independent of the reconstruction $r_{\boldsymbol{\theta}}$. In \cite{baldassarre,gozcu}, it is shown that the reconstruction error can be reduced by taking into account the reconstruction process while subsampling. For example, the CRLB cost can be replaced by mean-squared error or peak signal to noise ratio which is computed after estimating the sparse vector by a fixed reconstruction $r_{\boldsymbol{\theta}}$.
Details of this modified greedy algorithm that depends on a reconstruction method is summarized in Algorithm~\ref{algo:greedy_modified}. In Algorithm~\ref{algo:greedy_modified}, the sample locations selected at the end of the $k$-th iteration are denoted as $\mathcal{K}^{(k)}$. At the $k$-th iteration, to evaluate [S1], we need to apply a reconstruction algorithm $(N-k+1)$-times. This results in an overall of $KN- \frac{K(K-1)}{2}$ recalls of the reconstruction algorithm. 
	
\begin{algorithm}[!h]
	\caption{Modified Greedy Algorithm}
	\label{algo:greedy_modified}
	\begin{algorithmic}
	\State {\bf Inputs:} Data $\mathcal{D}$, full-sample indices $\mathcal{N}$, and a reconstruction method $r_{\boldsymbol{\theta}}$
	\State {\bf Output:} Sampling set $\mathcal{K}$
	\State {\bf Initialize:} $\mathcal{K} =\emptyset$
	\For{$k=1$ to $K$ } 
	
	\State[S1] For $i \in \mathcal{N}\backslash \mathcal{K}$, estimate $\mathbf{x}_{q, \mathcal{K} \cup \{i\}} = r_{\boldsymbol{\theta}} \left(\text{diag}(\mathbf{c})\mathbf{f}_q\right)$ where $c_k = 1$ for $k \in \mathcal{K} \cup \{i\}$ and the rest of the coefficients are zero
	\State[S2]   $i ^{*}  = \displaystyle \underset{i \in \mathcal{N}\backslash \mathcal{K}}{\min} \quad \frac{1}{Q} \sum_{q=1}^Q \mathrm{C}\left(\mathbf{x}_q, \mathbf{x}_{q, \mathcal{K} \cup \{i\}} \right)$  \\
	\State[S3] $\mathcal{K} = \mathcal{K} \cup \{i^*\}$ and $\mathcal{K}^{(k)} = \mathcal{K}$
	\EndFor
	\end{algorithmic}
	\end{algorithm}

	In Algorithm~\ref{algo:greedy_modified}, one starts with zero samples and sequentially adds newer samples. This approach may not be efficient, especially in the presence of noise, due to fewer measurements/samples to begin with compared with the number of unknowns. An alternative approach is to start with full samples and sequentially remove one sample at a time until the desired number of samples are retained. The sample removed at each iteration is the one for which a change in the cost function is smallest as described in Algorithm~\ref{algo:greedy_modified2}. 

	\begin{algorithm}[!t]
		\caption{Modified Greedy Algorithm - 2}
		\label{algo:greedy_modified2}
		\begin{algorithmic}
			\State {\bf Inputs:} Data $\mathcal{D}$, full sample indices $\mathcal{N}$, and a reconstruction method $r_{\boldsymbol{\theta}}$
			\State {\bf Output:} Sampling set $\mathcal{K} \subseteq \mathcal{N}$ with $|\mathcal{K}| = K$
			\State {\bf Initialize:} $\mathcal{K} =\mathcal{N}$
			\For{$k=1$ to $N-K$ } 
			\State[S1] For $i \in \mathcal{K}$, estimate $\mathbf{x}_{q, \mathcal{K} \backslash \{i\}} = r_{\boldsymbol{\theta}} \left(\text{diag}(\mathbf{c})\mathbf{f}_q\right)$ where $c_k = 1$ for $k \in \mathcal{K} \backslash \{i\}$ and the rest of the coefficients are zero
			\State[S2]   $i ^{*}  = \displaystyle \underset{i \in  \mathcal{K}}{\min} \quad \frac{1}{Q} \sum_{q=1}^Q \mathrm{C}\left(\mathbf{x}_q, \mathbf{x}_{q, \mathcal{K} \backslash \{i\}} \right)$  \\
			\State[S3] $\mathcal{K} = \mathcal{K} \backslash \{i^*\}$ and $\mathcal{K}^{(k)} = \mathcal{K}$
			\EndFor
		\end{algorithmic}
	\end{algorithm}
	This algorithm requires $\frac{(N-K)(N+K+1)}{2}$ recalls of the reconstruction method. Algorithm~\ref{algo:greedy_modified} requires fewer computations compared to that of Algorithm~\ref{algo:greedy_modified2} for $K<\frac{N}{2}$ and vice versa. The sampling pattern and hence the reconstruction accuracy of the two algorithms depend on the data used and noise levels. In addition, the cost functions are not necessarily sub-modular and optimality of the algorithms is not guaranteed.  
	
	
	\subsection{LISTA-Based Reconstruction}
	In our approach, we use LISTA as our recovery strategy which is discussed next. To estimate a sparse $\mathbf{x}$ from its compressed measurements $\mathbf{\bar{f}} = \mathbf{B}\mathbf{x}$, where $\mathbf{B} = \text{diag}(\mathbf{c}) \text{diag}(\mathbf{h}) \mathbf{A}$ (cf. \eqref{eq:fri_fourier2}) an $l_1$-relaxed optimization problem is considered
	\begin{align}
	\underset{\mathbf{x} \in \mathbb{C}^N }{\min}\quad 	\frac{1}{2} \|\mathbf{\bar{f}} - \mathbf{B} \mathbf{x}\|_2^2 + \bar{\lambda} \|\mathbf{x}\|_1,
	\label{eq:l1}
	\end{align}
	where $\bar{\lambda} >0$ is a sparsity enforcing regularization parameter. One of the ways to solve \eqref{eq:l1} is to apply ISTA which starts from an initial solution $\mathbf{x}^{(0)}$ and iteratively updates it by using the following step:
	\begin{align}
	\mathbf{x}^{(i+1)} = \mathcal{T}_{\frac{\bar{\lambda}}{\mu}} \left\{ \left(  \mathbf{I}_N - \frac{1}{\mu}\mathbf{B}^{\mathrm{H}} \mathbf{B} \right) \mathbf{x}^{(i)} + \frac{1}{\mu} \mathbf{B}^{\mathrm{H}} \mathbf{\bar{f}} \right \}, i = 1, 2, \cdots, 
	\label{eq:ista_step}
	\end{align}
	where $\mu$ is a constant parameter that controls the step size of each iteration and $\mathbf{I}_N$ denotes an $N \times N$ identity matrix. In \eqref{eq:ista_step}, $\mathcal{T}_{\alpha}\{\cdot \}$ is an elementwise soft-thresholding operator defined as $\mathcal{T}_{\alpha}\{x \} = \text{sign}(x) \, \max\{|x|, 0 \}$. 
	
	\begin{figure}
		\centering
		\includegraphics[width= 3.5in]{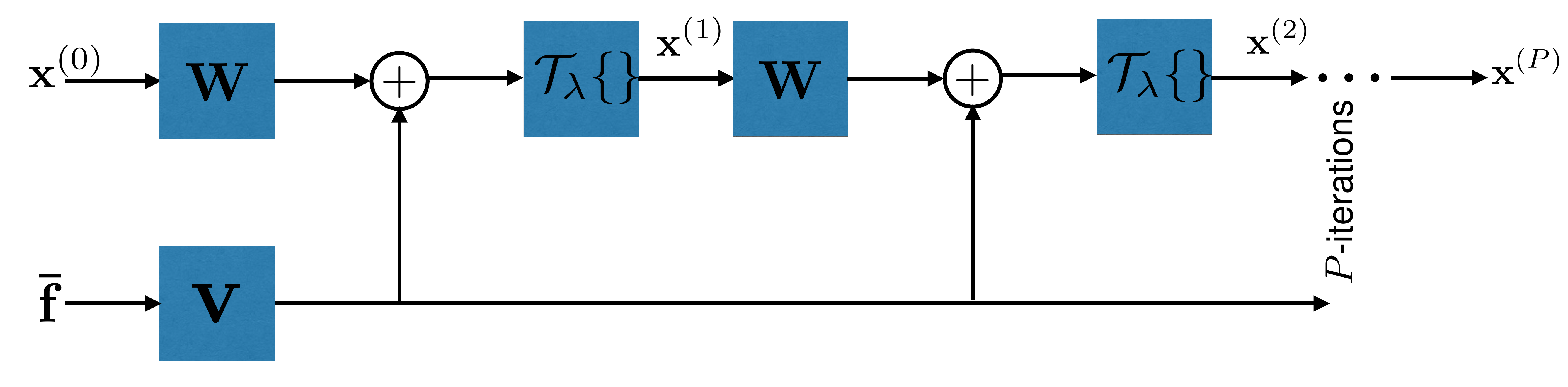}
		\caption{Flow diagram of ISTA and network architecture for LISTA: While the set of parameters $\boldsymbol{\theta} = \{\lambda, \mathbf{W}, \mathbf{V}\}$  is fixed in ISTA whereas in LISTA $\boldsymbol{\theta}$ is learned for a given set examples.  }
		\label{fig:ista}
	\end{figure}
	
	By replacing $\lambda = \frac{\bar{\lambda}}{\mu}$, $\mathbf{W} =\left(  \mathbf{I}_N - \frac{1}{\mu}\mathbf{B}^{\mathrm{H}} \mathbf{B} \right)$, and $\mathbf{V} =  \frac{1}{\mu} \mathbf{B}^{\mathrm{H}}$, in \eqref{eq:ista_step}, we have 
	\begin{align}
	\mathbf{x}^{i+1} = \mathcal{T}_{\lambda} \left\{ \mathbf{W} \mathbf{x}^i + \mathbf{V} \mathbf{\bar{f}} \right \}.
	\label{eq:lista_step}
	\end{align}
	The iterations of ISTA are summarized in Fig.~\ref{fig:ista}.
	Note that $\boldsymbol{\theta} = \{\lambda, \mathbf{W}, \mathbf{V}\}$ are reconstruction parameters. While  $\boldsymbol{\theta}$ is fixed in ISTA, it is learned from the examples using unrolled iterations as shown in Fig.~\ref{fig:ista}. LISTA maps each iteration into a layer of the network. By stacking a finite number of layers (for example $P$ layers in Fig~\ref{fig:ista}), a network is constructed and by using end to end training $\boldsymbol{\theta}$ is learned from a set of examples to minimize $\displaystyle \sum_{q=1}^Q \|\mathbf{x}_q - \mathbf{\hat{x}}_q \|_2^2$ where $\mathbf{\hat{x}}_q$ is the output of the $P$-th layer.
	
	The LISTA-based reconstruction algorithm is well suited for our goal; it leads to a reconstruction which is parametric, data-driven, and does not require exact knowledge of the measurement matrix $\mathbf{B}$ which is a function of $\mathbf{h}$. 
	
	%
	%
	\begin{algorithm}[!t]
		\caption{Joint Subsampling and Recovery Algorithm}
		\label{algo:jsr}
		\begin{algorithmic}
			\State {\bf Inputs:} Data $\mathcal{D}$ and full sample indices $\mathcal{N}$
			\State {\bf Initialize:} $\mathcal{K}^{(0)} =\emptyset$
			\For{$k=1$ to $K$ } 
			
			\State[S1] {\bf for all} $i \in \mathcal{N}\backslash \mathcal{K}^{(k-1)}$ {\bf do}
			\\ \qquad  \qquad 
			\begin{minipage}{20 em}
				$\mathbf{(a)}$ For a binary-valued vector $\mathbf{c}_i \in \{0, 1\}^{|\mathcal{N}|}$, set $\displaystyle \text{supp}\{\mathbf{c}_i\} = \mathcal{K}^{(k-1)} \cup \{i\}$
			\end{minipage}\\
			\\ \qquad  \qquad
			\begin{minipage}{20 em}
				$\mathbf{(b)}$ $\displaystyle \boldsymbol{\theta}^{(k)}_i = \underset{\boldsymbol{\theta}}{\arg \min} \frac{1}{Q} \sum_{q=1}^Q  \| \mathbf{x}_q -   r_{\boldsymbol{\theta}} \left(\text{diag}(\mathbf{c}_i)\mathbf{f}_q \right) \|_2^2$ where $r_{\boldsymbol{\theta}}$ is a LISTA-based reconstruction.
			\end{minipage} \\
			\\ \qquad  \qquad 
			\\ \qquad  \qquad 
			\begin{minipage}{20 em}
				$\mathbf{(c)}$ $\mathbf{x}_{q, i} = r_{\boldsymbol{\theta}^{(k)}_i} \left(\text{diag}(\mathbf{c}_i)\mathbf{f}_q\right)$ for $q =1, \cdots, Q$
			\end{minipage}\\
			\qquad  \quad {\bf end for}
			\State[S2]   $i ^{(k)}_{*}  = \displaystyle \underset{i \in \mathcal{N}\backslash \mathcal{K}^{(k-1)}}{\arg\min} \quad \frac{1}{Q} \sum_{q=1}^Q \mathrm{C}\left(\mathbf{x}_q, \mathbf{x}_{q,i} \right)$  \\
			\State[S3] $\mathcal{K}^{(k)} = \mathcal{K}^{(k-1)} \cup \{i^{(k)}_{*} \}$
			\EndFor
			\State {\bf Output:} Optimal sampling set $\mathcal{K}^{K} \subseteq \mathcal{N}$ with $|\mathcal{K}^{K}| = K$ and corresponding reconstruction parameters $\boldsymbol{\theta}_{i^{(k)}_{*}}$
		\end{algorithmic}
	\end{algorithm}

	\begin{figure}
		\centering
		\subfigure[Unfolding of Algorithm~\ref{algo:jsr}]{\includegraphics[width= 3.5in]{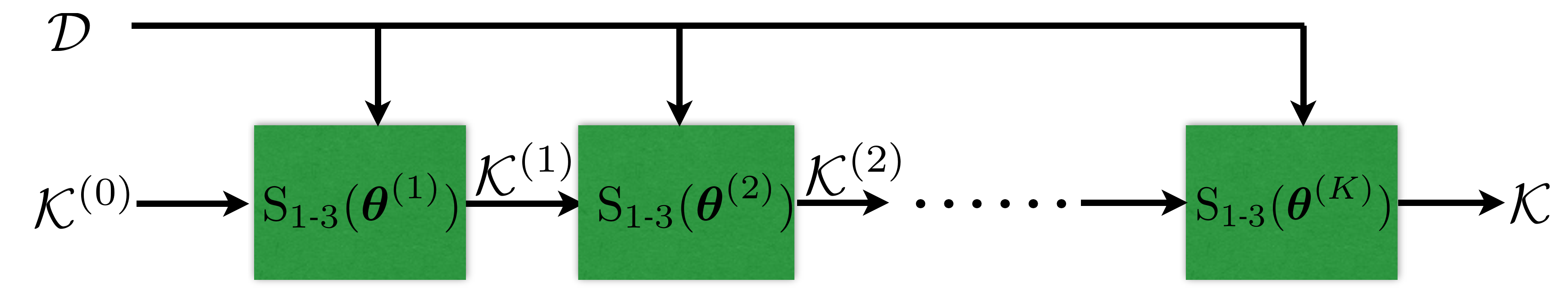}\label{fig:gu1}}\\
		\subfigure[Flow diagram of $\mathrm{S}_{1\text{-}3}(\boldsymbol{\theta})$ of $k$-th iteration]{\includegraphics[width= 3.5in]{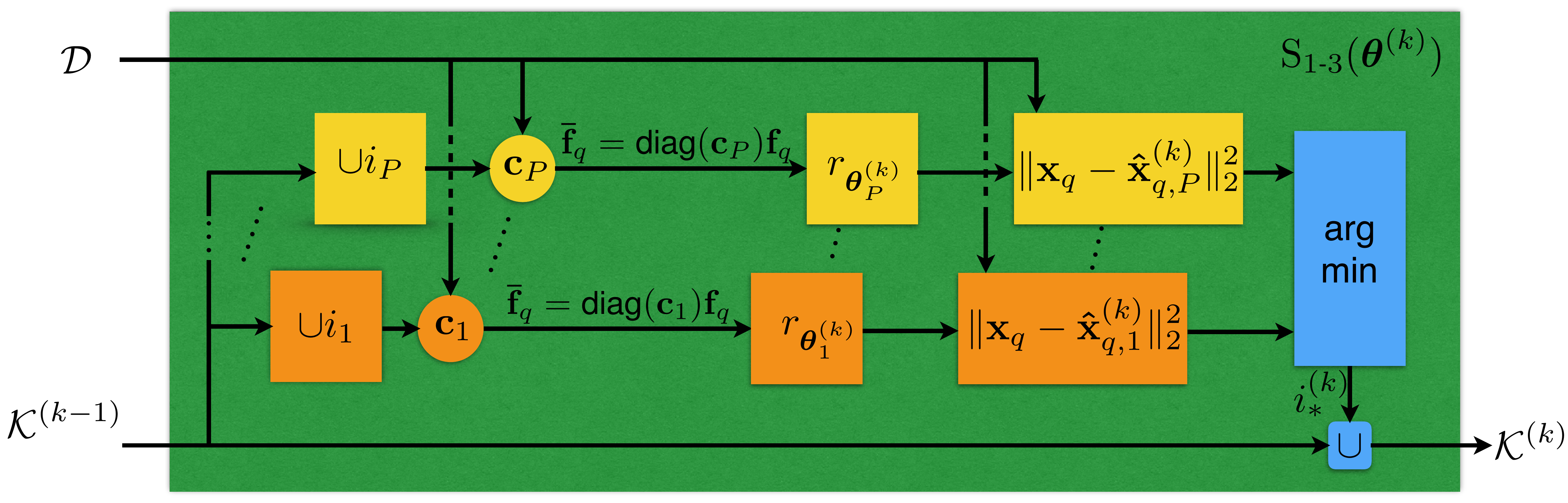}\label{fig:gu2}}
		\caption{A flow diagram of Algorithm~\ref{algo:jsr}: (a) Steps [S1]-[S3] are denoted by $\mathrm{S}_{1\text{-}3}(\boldsymbol{\theta})$, and the parameter $\boldsymbol{\theta}$ of the reconstruction algorithm varies for each iteration. (b) For $k$-th iteration, $P = N-k$; For each $r_{\boldsymbol{\theta}}$, LISTA network shown in Fig.~\ref{fig:ista} is used and the parameters are learned from the given data.}
		\label{fig:greedy_unroll}
	\end{figure}
	
	\subsection{Proposed Joint Sampling and Reconstruction (JSR) Algorithm}
	Both Algorithm~\ref{algo:greedy_modified} and its variant Algorithm~\ref{algo:greedy_modified2} are functions of the parameter of the reconstruction algorithm $\boldsymbol{\theta}$. While it is fixed in both of these algorithms, finding an optimal $\boldsymbol{\theta}$ for a given set of examples results in a solution to the joint optimization problem in \eqref{eq:opt_joint2}. To this end, we use the LISTA-based reconstruction method while the greedy algorithm determines the sampling set.
	Details of the proposed joint sub-sampling and reconstruction approaches are described in Algorithm~\ref{algo:jsr}. As in Algorithm~\ref{algo:greedy_modified}, we start with zero samples and sequentially add a new sample. Alternatively, one can also start from full samples and remove one at a time as in Algorithm~\ref{algo:greedy_modified2}. 
	
	Comparing Algorithm~\ref{algo:jsr} and Algorithm~\ref{algo:greedy_modified}, we note that most of the steps remain the same, except the step [S1] where we optimize the reconstruction parameters for each sampling pattern in  Algorithm~\ref{algo:jsr} compared to a fixed reconstruction method in Algorithm~\ref{algo:greedy_modified}. Since the sampling mechanism is non-parametric the proposed JSR algorithm is generic and can be optimized for any reconstruction method.

	A flow diagram of the JSR algorithm is shown in Fig.~\ref{fig:greedy_unroll} where for simplicity of the discussion the steps [S1]-[S3] are denoted by an operator $\mathrm{S}_{1\text{-}3}(\boldsymbol{\theta})$. At the $k$-th iteration, the operator $\mathrm{S}_{1\text{-}3}(\boldsymbol{\theta})$ acts on the previous sampling pattern $\mathcal{K}^{(k-1)}$ and outputs $\mathcal{K}^{(k)}$, that is, $\mathcal{K}^{(k)} = \mathrm{S}_{1\text{-}3}(\boldsymbol{\theta})\{\mathcal{K}^{(k-1)}\}$. Note that the parameters of the reconstruction algorithm vary for each iteration as the number of samples vary with iteration. Moreover, within each iteration, for different sampling patterns, different reconstruction parameters are used as shown in Fig.~\ref{fig:greedy_unroll}(b). 
	
	The optimization problem in step [S1](b) is solved by a deep-learning approach. Specifically, each reconstruction step is replaced by a LISTA network, as shown in Fig.~\ref{fig:ista}, and the parameters are learned. There is no backpropagation step from one iteration to another of [S1]. Because of which, there is no derivative step over the sampling patterns and the differentiability issue does not arise.
	
	Suppose the network is trained for $K$ samples. Then at the end of the $K$-th iteration of the greedy algorithm, we have a pair of optimal sampling patterns and corresponding reconstruction parameters $\{\mathcal{K}^{(k)}, \boldsymbol{\theta}_{i_{*}}^{(k)}\}$ for $k  = 1, \cdots, K$. Hence, if the goal is to choose fewer samples than $K$, then we already have a solution and no training is required. Whereas if it is required to increase the samples beyond $K$, then one can start from the selected the $K$ samples and follow the steps of Algorithm~\ref{algo:jsr} to add additional samples without starting from scratch. 
	
	\subsection{JSR Training}
	In Algorithm~\ref{algo:jsr}, only the sampling location set is passed between different sampling layers. Hence, we can use a simple sequential training approach to train each sampling layer (one iteration of the greedy algorithm). In particular, for each sampling layer, we trained a dedicated LISTA network for each candidate sampling pattern. Thus, we train all the LISTA networks within the same sampling layer in a highly-efficient parallel way. The  initial parameters for each LISTA network within the same sampling layer are randomly generated with the same seed, which ensures that the best sampling location can be fairly selected.
	The loss function corresponding to the $i$th sampling pattern in the $k$-th sampling layer is given by $\displaystyle \boldsymbol{\theta}^{(k)}_i = \underset{\boldsymbol{\theta}}{\arg \min} \frac{1}{Q} \sum_{q=1}^Q  \| \mathbf{x}_q -   r_{\boldsymbol{\theta}} \left(\text{diag}(\mathbf{c}_i)\mathbf{f}_q \right) \|_2^2$, which is used to adjust the trainable variables in LISTA. The Adam solver is used to stochastically optimize the network parameters, where the learning rate is set to be $0.001$. For the LISTA module, we set the number of unfolding layers to $10$. 
	
	In terms of computational complexity, the algorithm requires multiple training of the LISTA network. However, within each iteration of the greedy algorithm, the LISTA networks can be trained in parallel to reduce the training time. In applications such as MRI, where Fourier samples are typically considered in a two-dimensional plane, $N$, and $K$ are large. To reduce the computations, instead of adding one sample at a time as the current approach, one can sequentially add one sampling pattern from a set of predefined sampling patterns as in \cite{gozcu}. For example, one can sequentially add rows of samples from a two-dimensional Fourier plane.

	\section{Numerical Results}
	\label{sec:numerical_results}
	In this section, we provide numerical results to verify the performance of our proposed joint-sampling recovery scheme presented in Algorithm~\ref{algo:jsr}. To this end, we compared the following methods:\\
	{\bf Rand+FISTA:} The sub-sampling pattern is chosen randomly, and reconstruction is achieved by applying FISTA. \\
	{\bf G-CRLB+FISTA:} In this algorithm, the sub-sampling pattern is determined by using the greedy algorithm with a CRLB-based cost function given in \eqref{eq:crlb_cost}, and reconstruction is achieved by applying FISTA. A close form expression for CRLB for estimation of time-delays and amplitudes can be found in \cite{mulleti_fdoct}.  In the greedy algorithm, we started with full samples and sequentially removed one per iteration as in Algorithm~\ref{algo:greedy_modified2}.\\
	{\bf G-FISTA+FISTA:} We use Algorithm~\ref{algo:greedy_modified2} with FISTA as a recovery method in step [S1] to generate the sampling pattern and used FISTA for recovery. In principle, this algorithm is similar to that in \cite{baldassarre,gozcu}. In [S2], we used squared-error as a cost function: \begin{align}
	  \mathrm{C}\left(\mathbf{x}_q, \mathbf{x}_{q, \mathcal{K} \backslash \{i\}} \right) = \|\mathbf{x}_q- \mathbf{x}_{q, \mathcal{K} \backslash \{i\}}\|_2^2.
	  \label{eq:ls_cost}
	\end{align}
	We use the same cost function in the following three algorithms.
	\\
	{\bf G-FISTA+LISTA:} We use Algorithm~\ref{algo:greedy_modified2} with FISTA as recovery method in step [S1] to generate the sampling pattern. Then a LISTA network is optimized for recovery for the given sampling pattern. \\
	{\bf JSR-1} We apply joint-sampling and recovery Algorithm~\ref{algo:jsr} to determine the sampling pattern and recovery parameters.
	\\
	{\bf JSR-2}  Algorithm~\ref{algo:jsr} is modified to start from full samples and sequentially remove one sample per iteration as in Algorithm~\ref{algo:greedy_modified2}.\\
	Except {\bf JSR-1} and {\bf JSR-2} algorithms, all the algorithm mentioned above require the knowledge of the pulse shape $\mathbf{h}$.

	We consider two experiments. The first one is to compare different algorithms as mentioned above for a given data set. The second is to analyze the data-driven approach of the proposed algorithm. For both the experiments, the performance is evaluated in terms of NMSE in the estimation of time-delays and amplitudes of the FRI signals. Since both these quantities are encoded in the support and amplitudes of the sparse vector $\mathbf{x}$, respectively (cf. \eqref{eq:SoWe_grid}), we compute the NMSE for the test data $\{\mathbf{x}_i\}_{i=1}^{Q_{\text{Test}}}$ as follows:
	\begin{align}
	\text{NMSE} =   \frac{\displaystyle \sum_{i=1}^{Q_{\text{Test}}} \|\mathbf{x}_i-\hat{\mathbf{x}}_i\|_2^2}{\displaystyle \sum_{i=1}^{Q_{\text{Test}}} \|\mathbf{x}_i\|_2^2},
	\label{eq:nmse}
	\end{align}
	where $\hat{\mathbf{x}}_i$ is an estimate of $\mathbf{x}_i$. To determine accuracy for support recovery, we compute hit rate as 
	\begin{align}
	\text{Hit rate} = \frac{1}{L Q_{\text{Test}}} \sum_{i=1}^{Q_{\text{Test}}} \text{supp}(\mathbf{x}_i) \cap \text{supp}(\hat{\mathbf{x}}_i),
	\label{hit_rate}
	\end{align}
	where the set $\text{supp}(\mathbf{x}_i)$ denotes $L$ non-zero locations of $\mathbf{x}_i$. Hit rate equal to one implies that the support is estimated exactly over all the examples. To analyze the noisy scenario, we assume that the clean measurements $\mathbf{\bar{f}}$ are contaminated by circular, additive white Gaussian noise with zero mean and variance $\sigma^2$. For a given clean signal $\mathbf{\bar{f}}$, the SNR is computed as
	\begin{align}
	\text{SNR} =  \frac{\|\mathbf{\bar{f}}\|_2^2}{N \sigma^2}.
	\label{eq:snr}
	\end{align}

	\begin{figure} 
		\centering 
		\subfigure[Non-learning methods.]{ 
			\label{fig:subfig:conventional}
			\includegraphics[width=3in]{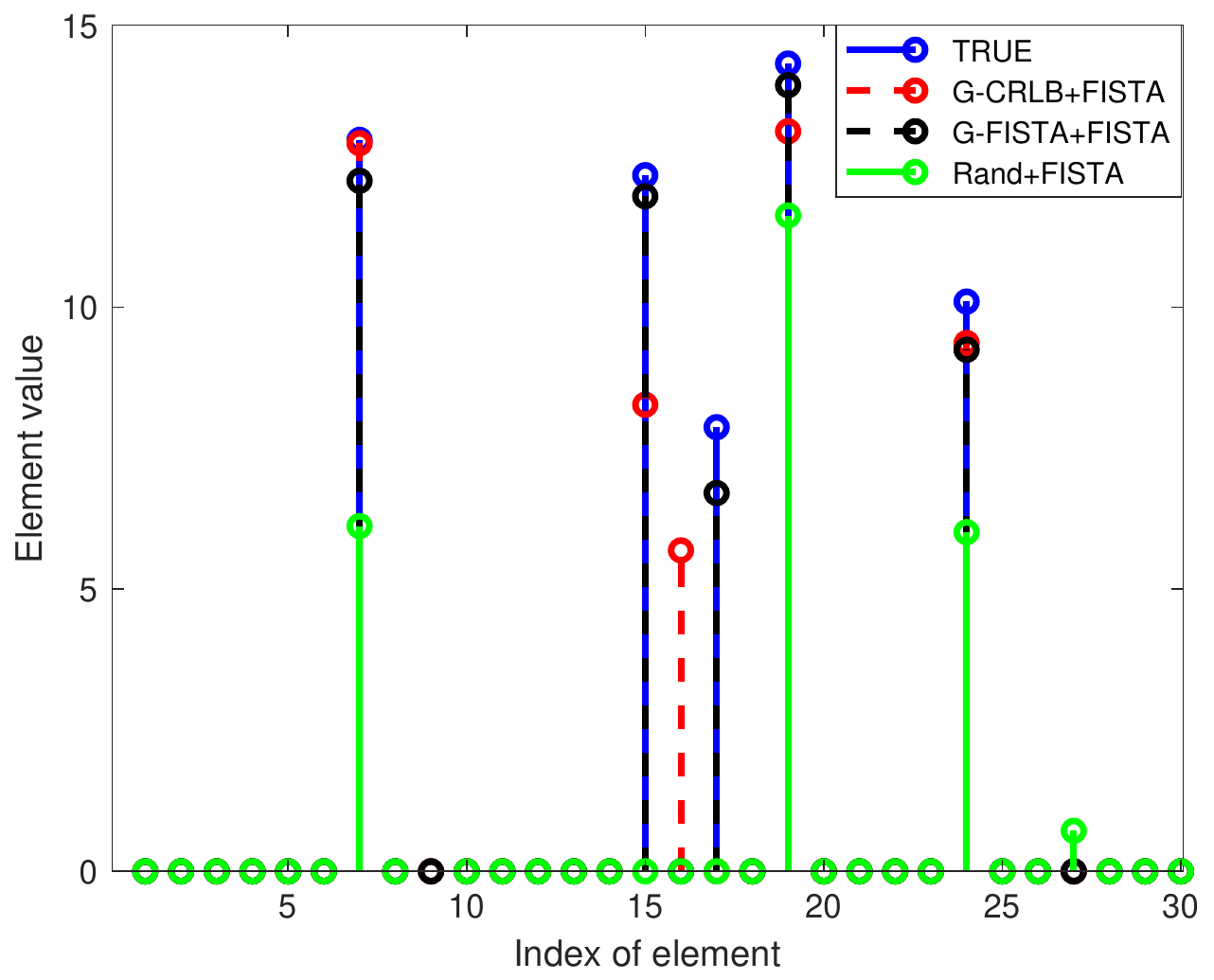} 
		} 
		\subfigure[Learning schemes]{ 
			\label{fig:subfig:learning} 
			\includegraphics[width=3in]{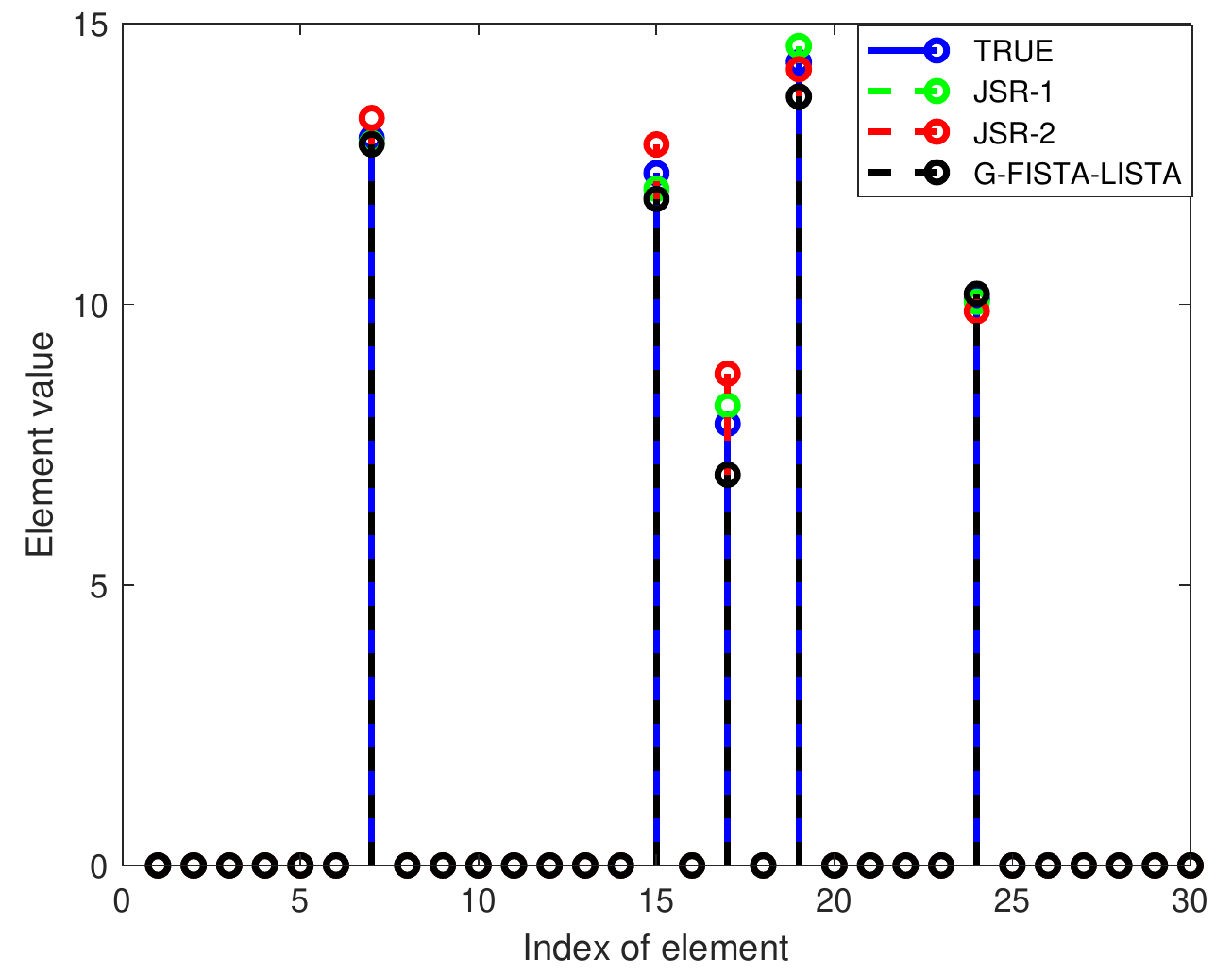}
		} 
		\caption{Estimation of the time-delays and amplitudes from sub-sampled Fourier measurements for different sampling and reconstruction schemes for $L=5$, $N=30$, $K=8$, and without noise case.} 
		\label{fig:K8} 
	\end{figure}

  	\begin{figure}[!h]
  	\centering
  	\includegraphics[width = 3.5 in]{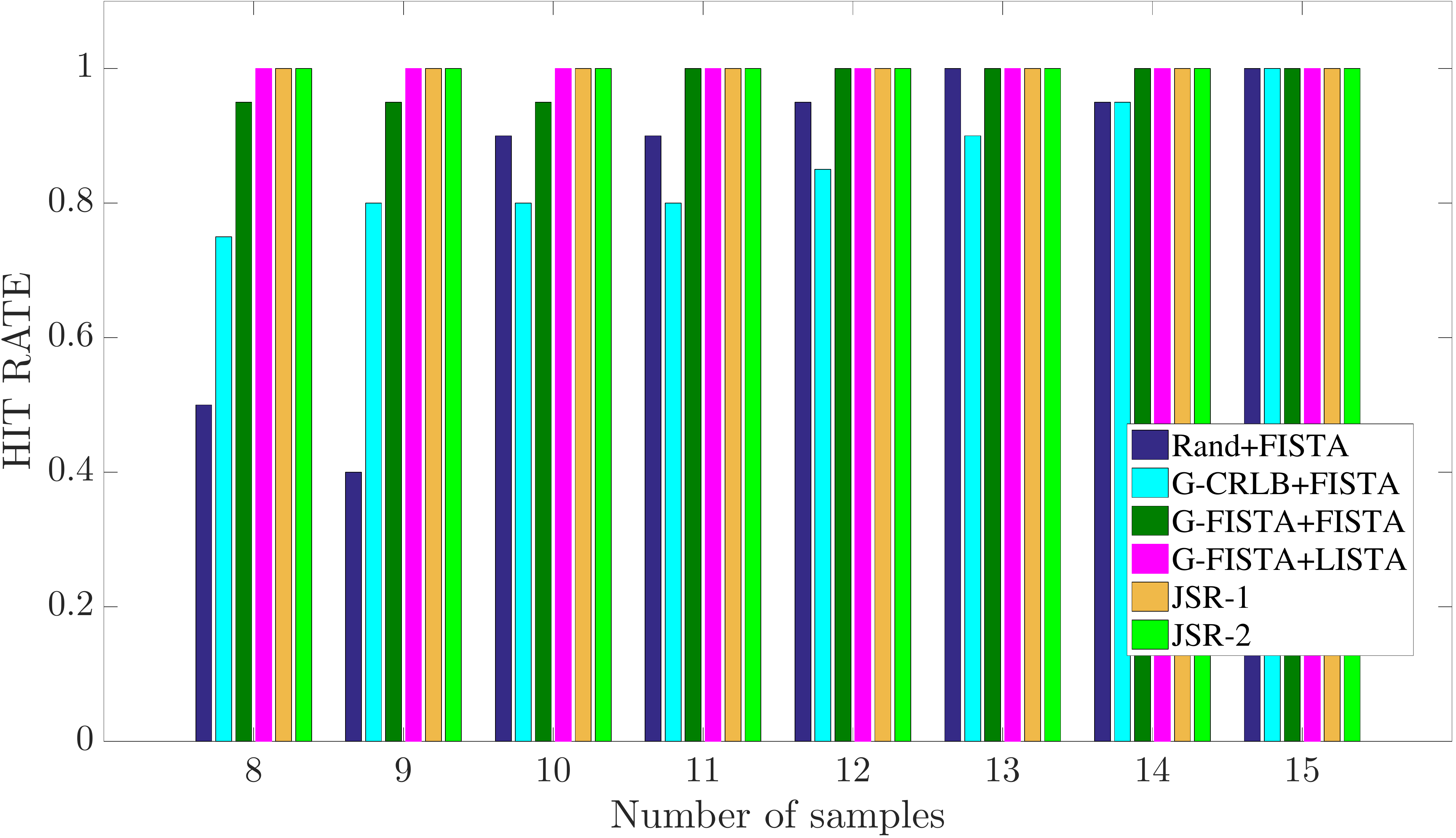}
  	\caption{A comparison of hit rates for different methods as a function of number of samples in the absence of noise. The learning-based approaches are able to determine the support perfectly for the sampling rate below the sub-Nyquist rate ($K <2L =10$). }
  	\label{fig:hitrate}
  \end{figure}
	
	\begin{figure} 
		\centering 
		\subfigure[Non-learning methods.]{ 
			\label{fig:subfig:conventional0}
			\includegraphics[width=3in]{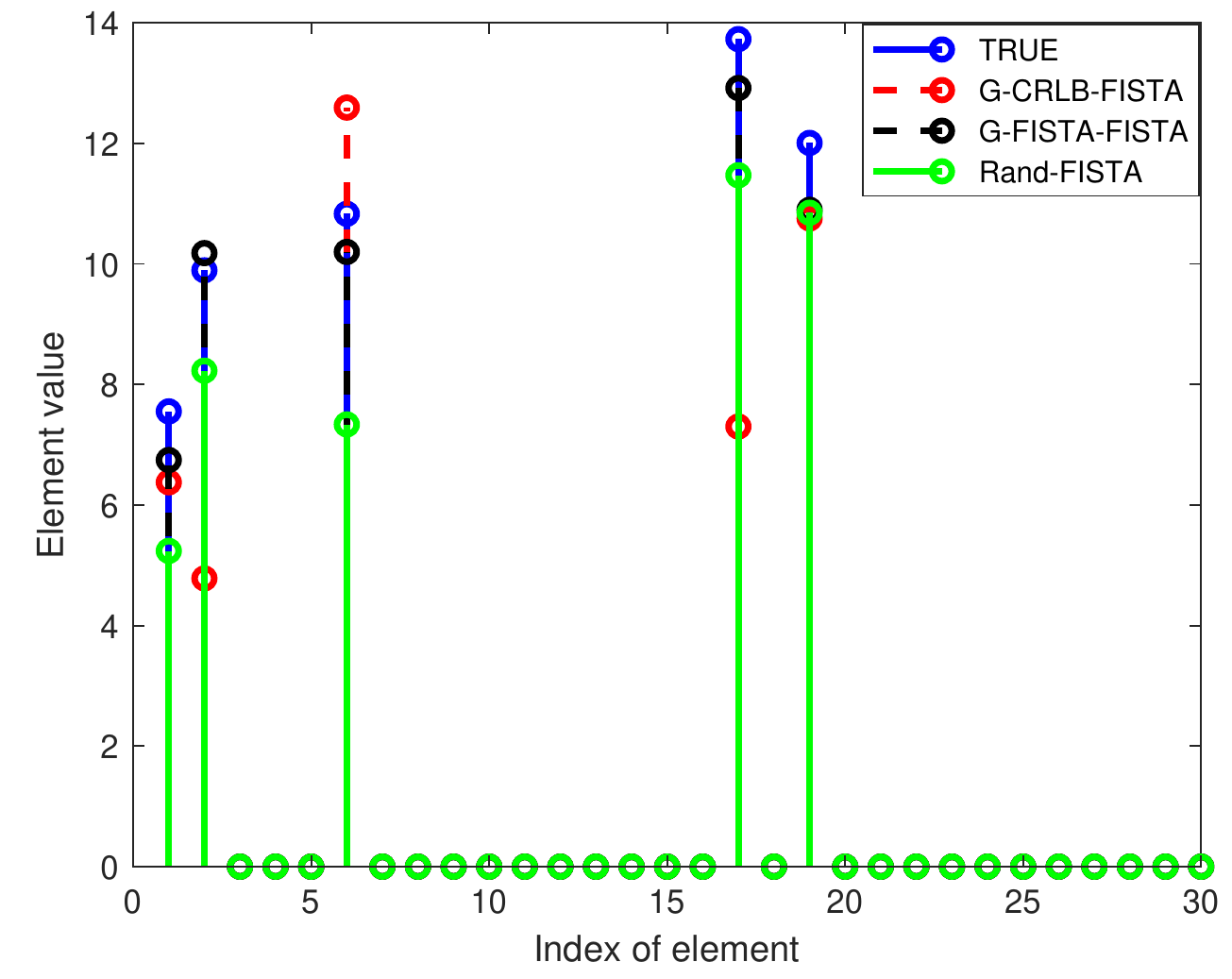} 
		} 
		\subfigure[Learning schemes]{ 
			\label{fig:subfig:learning0} 
			\includegraphics[width=3in]{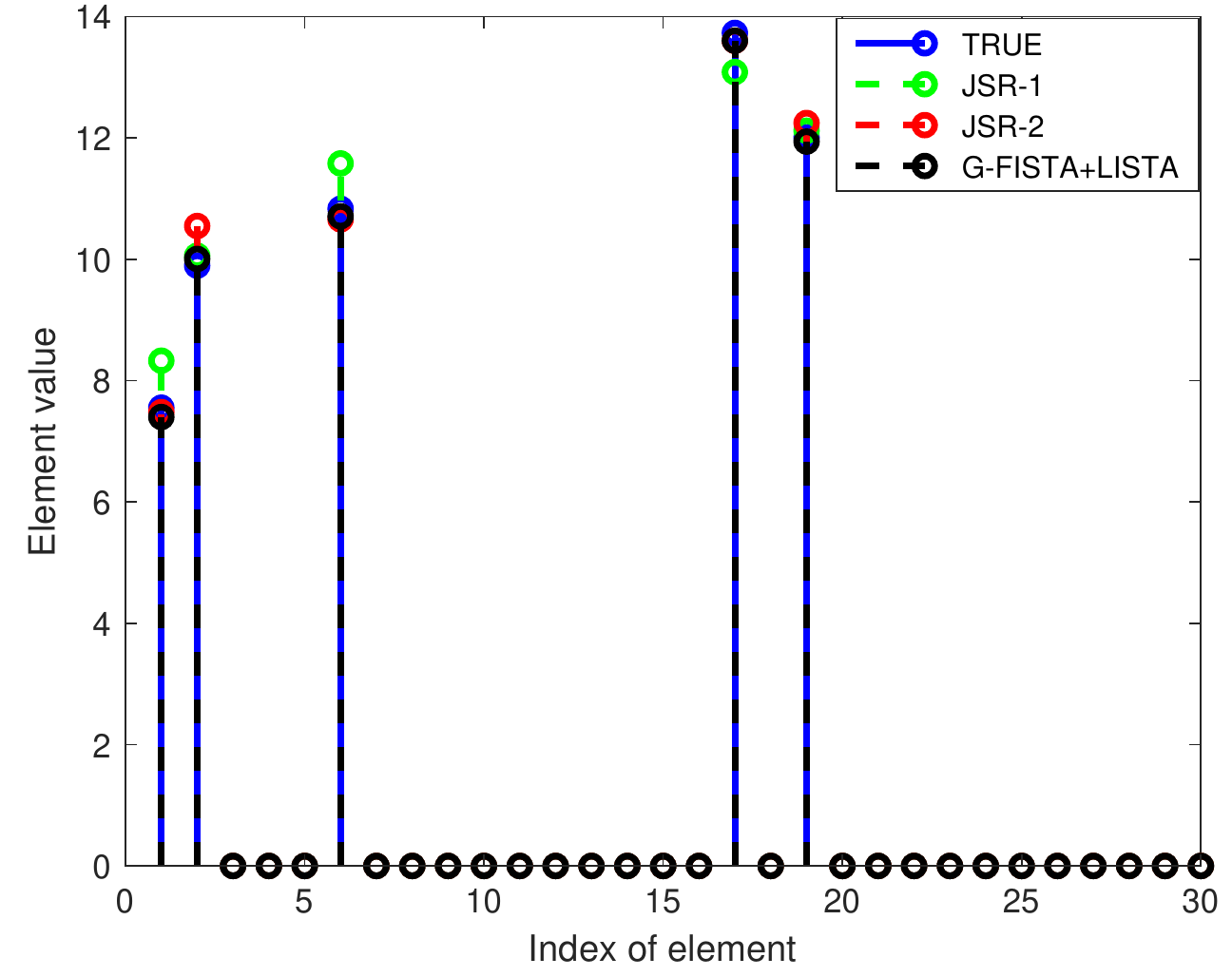}
		} 
		\caption{Estimation of the time-delays and amplitudes from sub-sampled Fourier measurements for different sampling and reconstruction schemes for $L=5$, $N=30$, $K=10$, and SNR$ = 30$ dB.} 
		\label{fig:Simu4} 
	\end{figure}

	\begin{figure}[!h]
		\centering
		\includegraphics[width = 3.0 in]{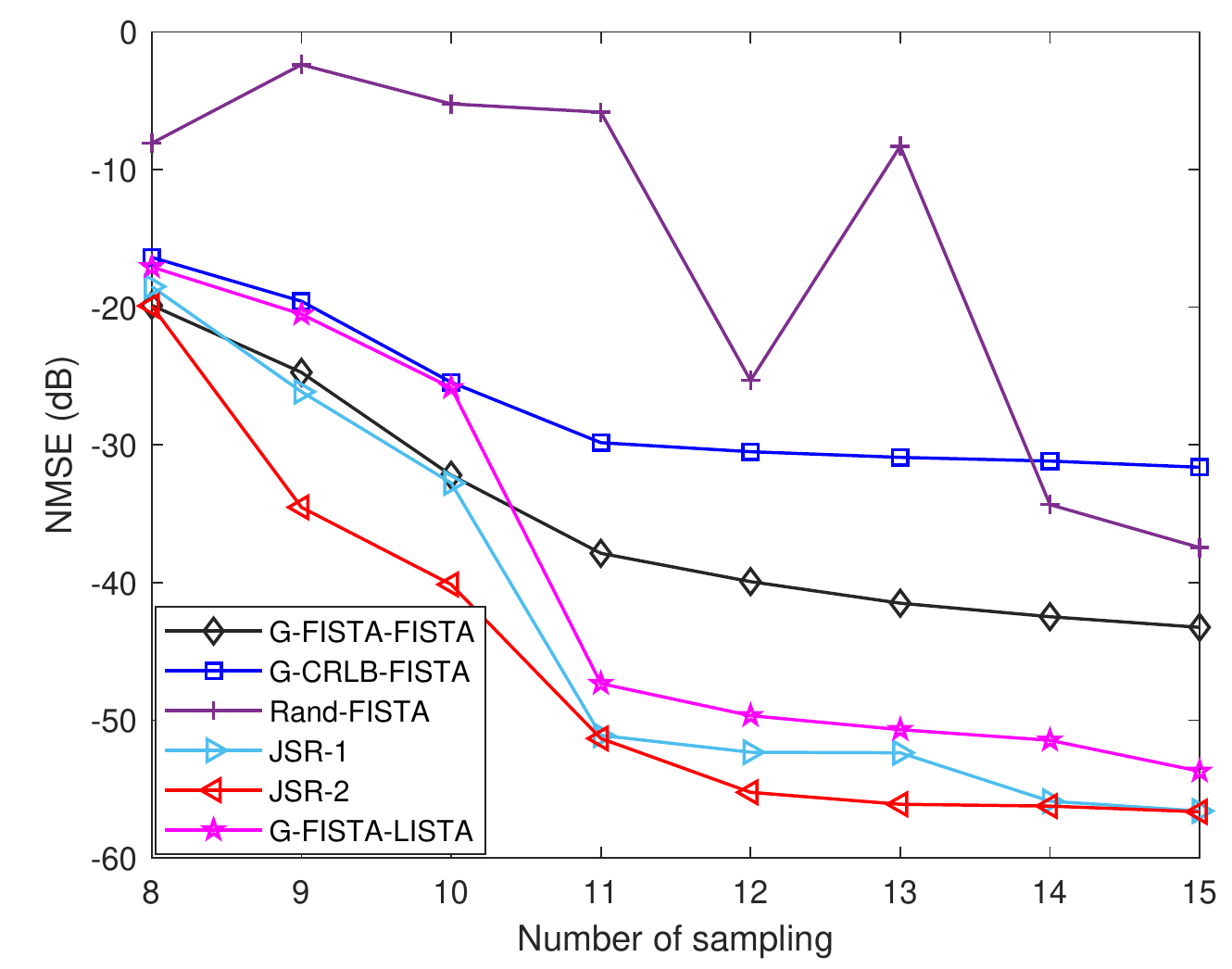}
		\caption{A comparison of performances different methods as a function of number of samples in the absence of noise. The proposed {\bf JSR-2} algorithm results in lowest NMSE among different methods.}
		\label{fig:WithoutNoise}
	\end{figure}

	\begin{figure}[!h]
		\centering
		\includegraphics[width = 3.0 in]{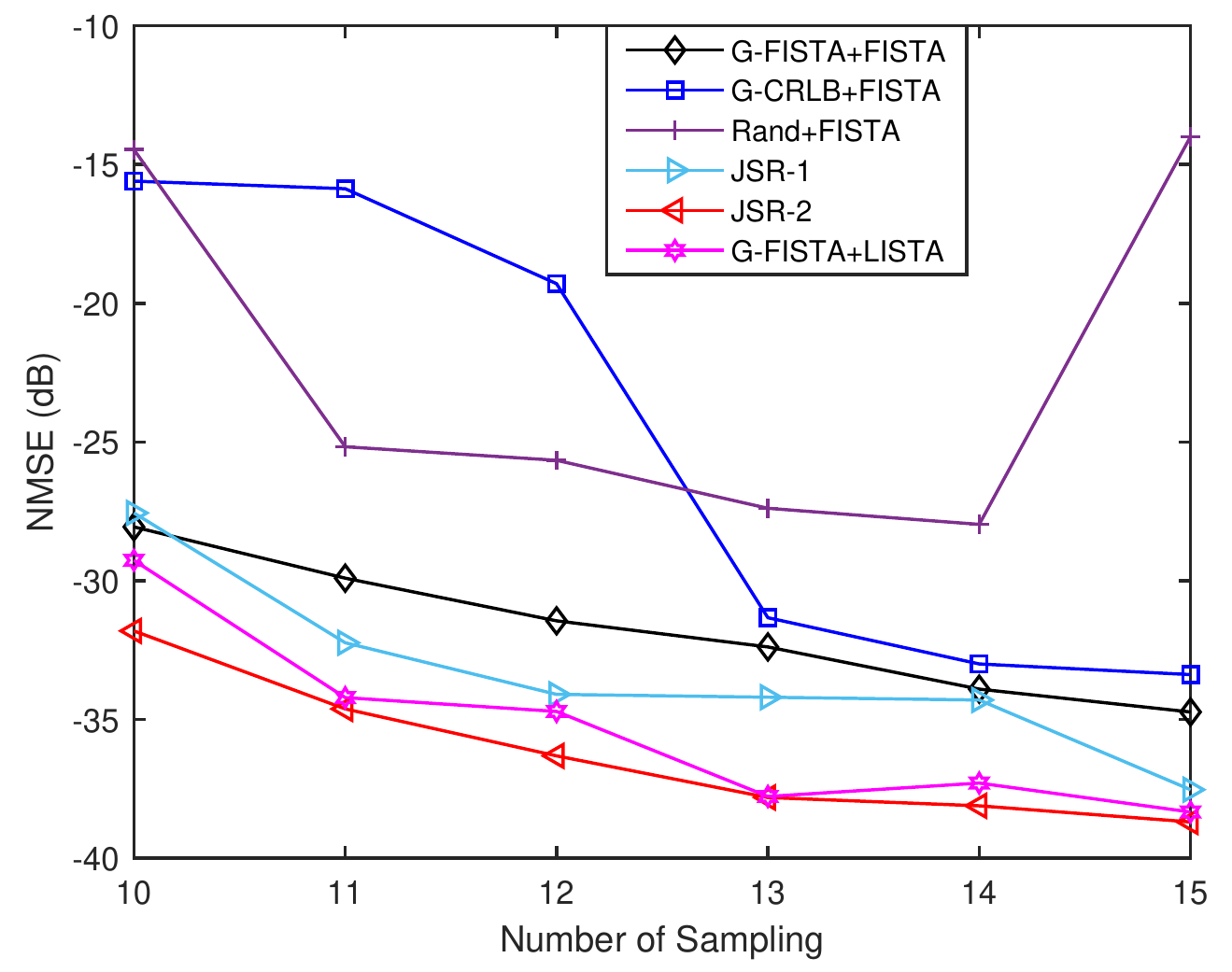}
		\caption{A comparison of performances different methods as a function of number of samples with $30$ dB SNR. The proposed {\bf JSR-2} algorithm results in lowest NMSE among different methods.}
		\label{fig:Simu1}
	\end{figure}

	\begin{figure}[!h]
		\centering
		\includegraphics[width = 3.0 in]{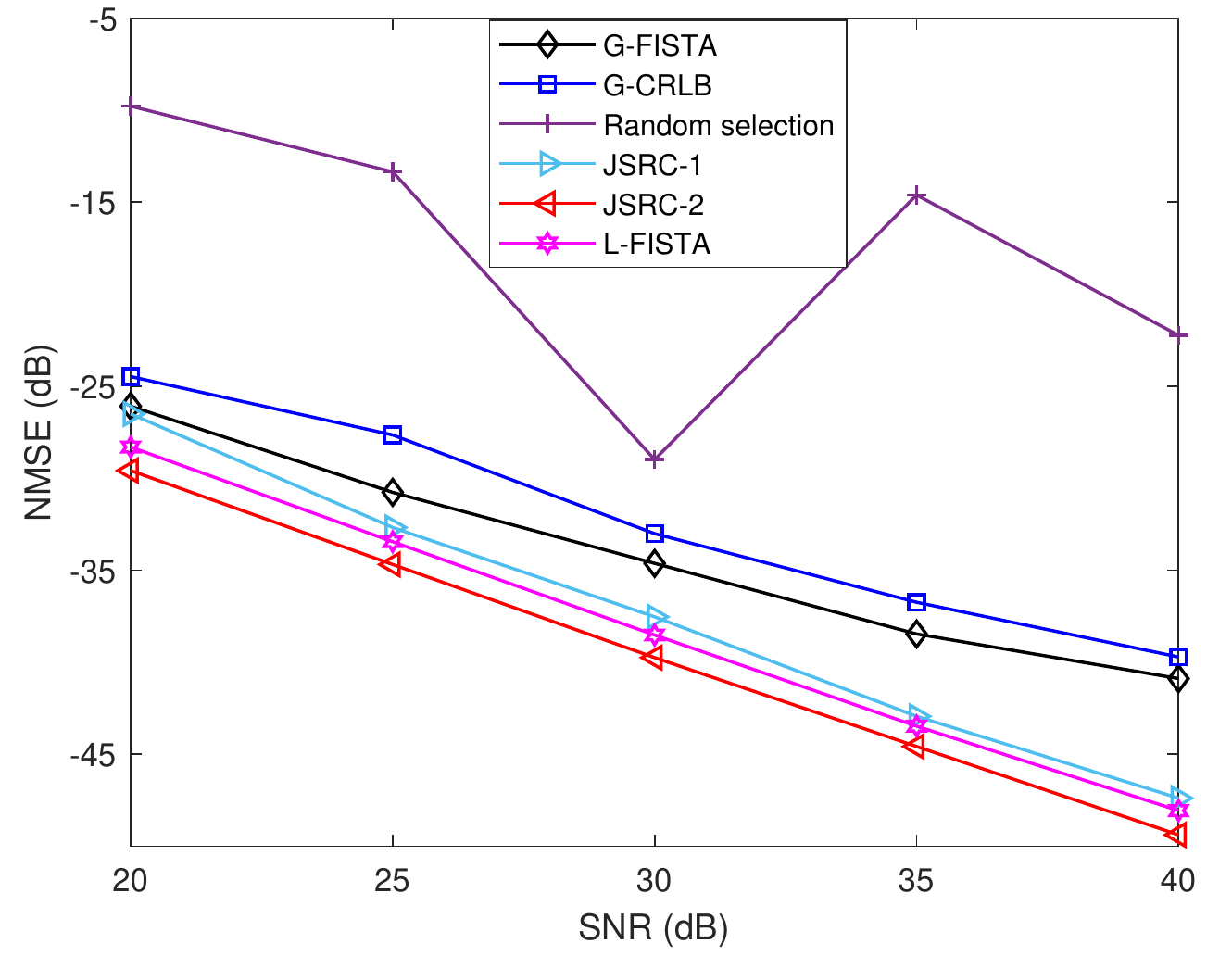}
		\caption{A Comparison of NMSE vs SNR for different methods for $K = 15$: learning-based methods have lower NMSE compared to non-learning-based algorithms {\bf JSR-2} algorithm results in lowest NMSE among different methods.  }
		\label{fig:Simu2}
	\end{figure}

	\subsection{Performance Comparison of Different Methods}
	\label{sec:sim_compare}
	In this experiment, we consider the problem of sampling and reconstructing FRI signals with $L=5$ and $t_{\max}$ = 1. To generate the training and test examples as in \eqref{eq:data}, we set $N = 30$. The support of $\mathbf{x}_q$ is generated uniformly at random over the set $\{1, 2, \cdots, N\}$ and amplitudes of its non-zero entries are i.i.d. Gaussian random variables with mean 10 and variance 3. We consider an FRI pulse $h(t)$ such that its Fourier samples, $\mathbf{h}$, are non-vanishing an non constant with significant variation. Specifically, the $n$-th sample of $\mathbf{h} \in \mathbb{C}^{N}$ is given as $0.01+ e^{-0.04(n-3 N/4)^2}+ e^{-0.04(n-N/8)^2}$. Note that large values of $\mathbf{h}$ may result in high SNR, and it may impact the selection sampling pattern. In addition, we used $Q=400,000$ examples for training, and $Q_{\text{ test}} = 5,000$ samples for testing. 
	
	For the above-mentioned experimental setup, we consider the problem of jointly optimizing the sampling kernel and reconstruction for $K = 8,  \cdots, 15$. First, we show results when the sampling rate is less than the sub-Nyquist rate. In particular, we consider $K = 8$ samples for recovery which is lower than the theoretical minimum of $2L = 10$ samples. One instance of recovery is shown in Fig.~\ref{fig:K8}. We observe that, except for {\bf Rand+FISTA} and {\bf G-CRLB+FISTA}, both learning-based methods and {\bf G-FISTA+FISTA} were able to estimate the time-delays of the FRI signals or support of the sparse vector accurately. The error is due to incorrect amplitude estimation. In Fig.~\ref{fig:hitrate}, we hit rate for different methods is plotted for different sampling rates. All the learning-based approaches are able to perfectly determine the support for all the sampling rates.	This leads to the conclusion that if the goal is to estimate only the time-delays as in the applications of radar imaging, one can go below the sub-Nyquist rates by carefully choosing the Fourier samples and reconstruction for a given data set.  
	
	In Fig. \ref{fig:Simu4}, we show an example of recovery of the sparse vector for SNR$ = 30$ dB and $K = 2L =10$. We observe that while all the methods were able to estimate the support or the time-delays of the FRI signals perfectly, amplitude estimates are closer to the true values in the learning-based methods (Fig.~\ref{fig:subfig:learning0}) compared to non-learning-based methods (Fig.~\ref{fig:subfig:conventional0}). For this particular example, among the learning-based methods, {\bf G-FISTA+FISTA} is able to estimate the amplitudes more accurately compared to {\bf JSR} methods. However, on average, the proposed {\bf JSR-2} algorithm has lowest NMSE for different values of $K$ and SNRs as discussed next.  
	
	In Fig.~\ref{fig:WithoutNoise} we compared the algorithms for different values of $K$ in the absence of noise. In comparison to {\bf JSR-1}, {\bf JSR-2} method has 8 dB lower NMSE for $K=10$ and the gap between NMSEs reduces as $K$ increases. While {\bf JSR-1} outperforms non-learning-based methods with more than 15 dB error margin, it improves upon {\bf G-FISTA+LISTA} by $5-15$ dB for different sampling rates. For $K = 8, 9$, except random sampling, all the methods results in  an NMSE in the range of $-15$ to $-20$ dB and {\bf JSR-2} has lowest error. In this particular simulation, we note that while $K=10$ is theoretical minimum sampling in the absence of noise, the practical algorithms may not results in perfect recovery at this rate.   
	
	To compare the performance of the methods in the presence of noise, in Fig. \ref{fig:Simu1} we show NMSE as a function of the number of samples for 30 dB SNR. The learning-based methods result in lower NMSE compared to non-learning-based algorithms. However, unlike the results in the absence of noise, {\bf JSR-1} method has higher NMSE compared to that of {G-FISTA+LISTA}. Moreover, the gap between NMSEs of {\bf JSR-1} and {\bf JSR-2} is larger. This behavior is a consequence of the fact that the latter approach starts with full measurements compared to zero measurements in the former approach. As a result, at any given iteration of the algorithm, {\bf JSR-2} method has a larger number of samples to estimate the sparse vectors compared to that in {\bf JSR-1} and hence, it is able to perform better.  
	
	In Fig. \ref{fig:Simu2}, we compare NMSEs of different methods for SNRs varying from 20 to 40 dB with $K=15$. We observe that the learning-based methods outperform non-learning-based methods with more than 5 dB improvement in NMSE, the gain in NMSE of {\bf JSR-1} compared to NMSE of either {\bf JSR-2} or {\bf G-FISTA+LISTA} is only $1-3$ dB. In the noisy scenario, {\bf G-FISTA+LISTA} algorithm is preferable over {\bf JSR-2} as one LISTA network has to be trained.      
	
\begin{figure} 
		\centering 
		\subfigure[Non-learning methods.]{ 
			\label{fig:subfig:conventional1}
			\includegraphics[width=3.0in]{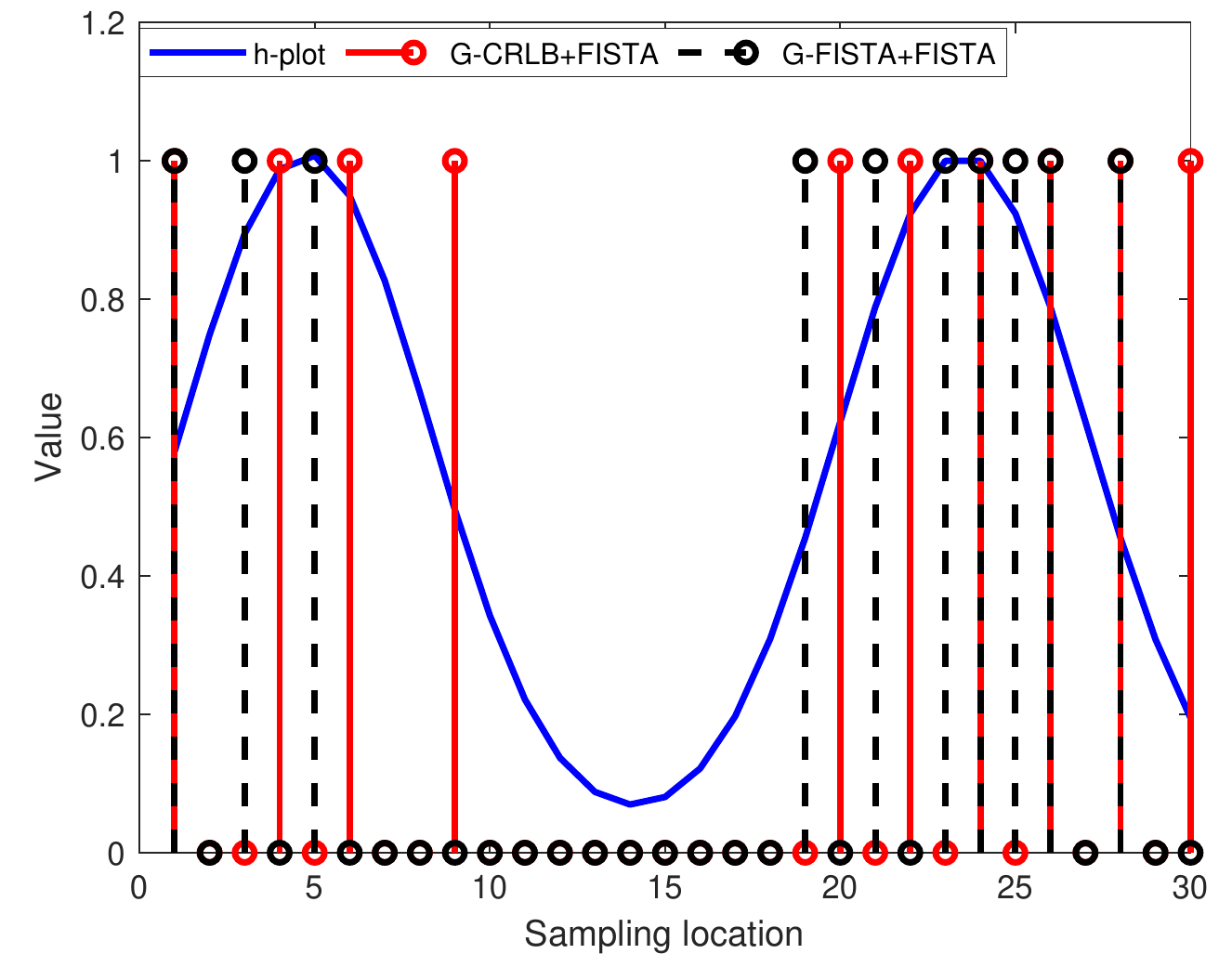} 
		} 
		\subfigure[Learning schemes]{ 
			\label{fig:subfig:learning1} 
			\includegraphics[width=3.0in]{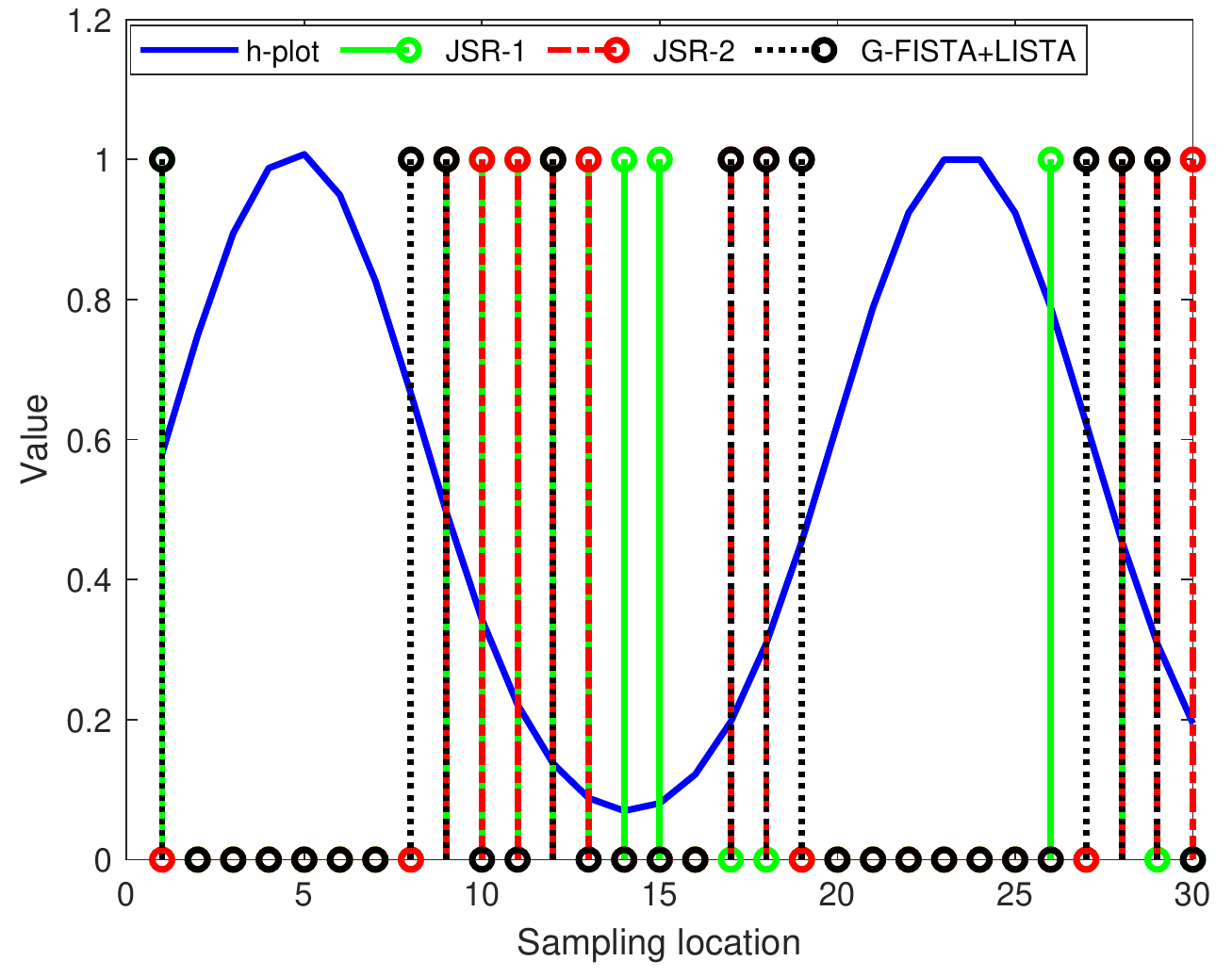}
		} 
		\caption{A comparison of sampling patterns estimated from various methods with a non-flat $\mathbf{h}$ (in blue). The sampling pattern of the learning-based methods are well spread compared to the non-learning algorithms.} 
		\label{fig:Simu5} 
	\end{figure}

	Next, we study the sampling patterns resulted from each method with respect to the frequency response of the pulse as shown in Fig. \ref{fig:Simu5} for $K=10$ and $\text{SNR} = 30$ dB. We observe that the sampling pattern of the learning-based methods s are well spread over the support of $\mathbf{h}$ compared to that of {\bf G-CRLB+FISTA} and {\bf G-FISTA+FISTA}. Both {\bf JSR} and {\bf G-FISTA+LISTA} depends on the recovery method, and we believe that a wider spread of the samples results in lower coherence in the resulting measurement matrix, which in turn leads to better estimation. Moreover, the sampling patterns show that it is not always necessary to consider samples from high-SNR regions, as shown in Fig.~\ref{fig:subfig:learning1}.

	\subsection{Data-Adaptability of {\bf JSR} Algorithm}
	\label{sec:data_adaptive}
	In this experiment, we study the data-adaptive nature of the proposed JSR algorithm to any additional structure within the data. We consider two different data sets for training and testing. For both the data sets, we have $N=30$ and $\mathbf{h}$ remains the same as in the previous experiment. The goal is to determine the sampling patterns and LISTA parameters $\{\mathcal{K}_1, \boldsymbol{\theta}_1\}$ and $\{\mathcal{K}_2, \boldsymbol{\theta}_2\}$ and then study the effect of mismatch between sampling pattern and recovery. For both the data sets, we assume $L=5$, but they differ in terms of their sparsity patterns. Specifically, for the first data set, the first two nonzero elements are chosen uniformly at random over the set $\{1, \cdots, 10\}$ and the remaining three from $\{21, \cdots, 30\}$. The nonzero values for the second data set are chosen uniformly at ransom over the set $\{11, \cdots, 20\}$. For both the data sets, the amplitude of the sparse vectors is chosen as in the previous experiment. We consider 40,000 samples for training and 2000 for testing with $|\mathcal{K}_1| = |\mathcal{K}_2 |= K$. In Table~\ref{tabel1}, we show NMSE incurred by {\bf JSR-2} algorithm for different combinations of sampling patterns and recovery parameters. We note that the error in the reconstruction is negligible when matched sampling and reconstructions are used, that is, for the combination $\{\mathcal{K}_1, \boldsymbol{\theta}_1\}$ and $\{\mathcal{K}_2, \boldsymbol{\theta}_2\}$. However, the NMSE is high when we used cross-combinations $\{\mathcal{K}_n, \boldsymbol{\theta}_m\}$, $n, m \in \{1, 2\}$ and $n \neq m$. We believe that the large NMSE for the cross-combinations is due to non-overlapping support of the sparse vectors. The observations lead to a conclusion that JSR approaches are able to better adapt to local structures of the sparse vectors. In addition, the learned sampling and reconstruction methods are not global and are highly data dependent.      
		\begin{table}\label{tabel1}
		\centering
		\caption{A comparison of NMSE while learning structured sparsity  }
		\begin{tabular}{ |c|c|c|c|c| } 
			\hline
			NMSE (dB) & $\mathcal{K}_1, {\boldsymbol{\theta}}_1 $ & $\mathcal{K}_2, {\boldsymbol{\theta}}_1$ & $\mathcal{K}_1, {\boldsymbol{\theta}}_2$ & $\mathcal{K}_2, {\boldsymbol{\theta}}_2$\\
			\hline
			Test Data 1 & -52.25  & -1.38  & 7.94 & -0.96 \\
			\hline
			Test Data 2 & 0.06 & 0.18 & 1.39 & -47.09 \\
			\hline
		\end{tabular}
	\end{table}
	
	\section{Conclusions}
	\label{sec:dis_conclusion}
	We propose a learning-based solution for a joint subsampling and reconstruction problem. The proposed algorithm is data-adaptive, circumvents the differentiability issue, and is flexible to change in the sampling rate. We show that the proposed algorithm can be utilized to design sampling kernels for FRI signals. Apart from sampling and reconstruction of FRI signals, our framework can be used for selecting $k$-space sampling in MRI, sinograms in computed tomography, and similar applications. Our methods have lower NMSE compared to independent design approaches. In addition, the sampling rate can be reduced below the sub-Nyquist rate which results in low-cost receivers.

	%
	%
	\bibliographystyle{IEEEtran}
	\bibliography{refs,los_refs,sparse_array_refs}

\begin{thebibliography}{10}
\providecommand{\url}[1]{#1}
\csname url@samestyle\endcsname
\providecommand{\newblock}{\relax}
\providecommand{\bibinfo}[2]{#2}
\providecommand{\BIBentrySTDinterwordspacing}{\spaceskip=0pt\relax}
\providecommand{\BIBentryALTinterwordstretchfactor}{4}
\providecommand{\BIBentryALTinterwordspacing}{\spaceskip=\fontdimen2\font plus
\BIBentryALTinterwordstretchfactor\fontdimen3\font minus
  \fontdimen4\font\relax}
\providecommand{\BIBforeignlanguage}[2]{{%
\expandafter\ifx\csname l@#1\endcsname\relax
\typeout{** WARNING: IEEEtran.bst: No hyphenation pattern has been}%
\typeout{** loaded for the language `#1'. Using the pattern for}%
\typeout{** the default language instead.}%
\else
\language=\csname l@#1\endcsname
\fi
#2}}
\providecommand{\BIBdecl}{\relax}
\BIBdecl

\bibitem{eldar_2015sampling}
Y.~C. Eldar, \emph{Sampling Theory: Beyond Bandlimited Systems}.\hskip 1em plus
  0.5em minus 0.4em\relax Cambridge University Press, 2015.

\bibitem{vetterli}
M.~Vetterli, P.~Marziliano, and T.~Blu, ``Sampling signals with finite rate of
  innovation,'' \emph{IEEE Trans. Signal Process.}, vol.~50, no.~6, pp.
  1417--1428, Jun. 2002.

\bibitem{bar_radar}
O.~Bar-Ilan and Y.~C. Eldar, ``Sub-{N}yquist radar via {D}oppler focusing,''
  \emph{IEEE Trans. Signal Process.}, vol.~62, no.~7, pp. 1796--1811, Apr.
  2014.

\bibitem{bajwa_radar}
W.~U. Bajwa, K.~Gedalyahu, and Y.~C. Eldar, ``Identification of parametric
  underspread linear systems and super-resolution radar,'' \emph{IEEE Trans.
  Signal Process.}, vol.~59, no.~6, pp. 2548--2561, Jun. 2011.

\bibitem{mulleti_icip}
S.~Mulleti, S.~Nagesh, R.~Langoju, A.~Patil, and C.~S. Seelamantula,
  ``Ultrasound image reconstruction using the finite-rate-of-innovation
  principle,'' in \emph{Proc. IEEE Int. Conf. Image Process. (ICIP)}, Oct.
  2014, pp. 1728--1732.

\bibitem{eldar_sos}
R.~Tur, Y.~C. Eldar, and Z.~Friedman, ``Innovation rate sampling of pulse
  streams with application to ultrasound imaging,'' \emph{IEEE Trans. Signal
  Process.}, vol.~59, no.~4, pp. 1827--1842, Apr. 2011.

\bibitem{fri_strang}
P.~L. Dragotti, M.~Vetterli, and T.~Blu, ``Sampling moments and reconstructing
  signals of finite rate of innovation: Shannon meets {S}trang-{F}ix,''
  \emph{IEEE Trans. Signal Process.}, vol.~55, no.~5, pp. 1741--1757, May 2007.

\bibitem{mulleti_paley}
S.~Mulleti and C.~S. Seelamantula, ``Paley-{W}iener characterization of kernels
  for finite-rate-of-innovation sampling,'' \emph{IEEE Trans. Signal Process.},
  vol.~65, no.~22, pp. 5860--5872, 2017.

\bibitem{stoica}
P.~Stoica and R.~L. Moses, \emph{Introduction to Spectral Analysis}.\hskip 1em
  plus 0.5em minus 0.4em\relax Upper Saddle River, NJ: Prentice Hall, 1997.

\bibitem{eldar_cs_book}
Y.~C. Eldar and G.~Kutyniok, \emph{Compressed Sensing: Theory and
  Applications}.\hskip 1em plus 0.5em minus 0.4em\relax Cambridge University
  Press, 2012.

\bibitem{omp}
S.~Mallat and Z.~Zhang, ``Matching pursuits with time-frequency dictionaries,''
  \emph{IEEE Trans. Signal Process.}, vol.~41, no.~12, p. 3397–3415, 1993.

\bibitem{ista}
I.~Daubechies, M.~Defrise, and C.~D. Mol, ``An iterative thresholding algorithm
  for linear inverse problems with a sparsity constraint,'' \emph{Comm. on Pure
  and Applied Mathematics}, vol.~57, pp. 1413--1457, 2004.

\bibitem{eldar_optimization_book}
D.~P. Palomar and Y.~C. Eldar, \emph{Convex Optimization in Signal Processing
  and Communications}.\hskip 1em plus 0.5em minus 0.4em\relax Cambridge
  University Press, 2010.

\bibitem{fista}
A.~{Beck} and M.~{Teboulle}, ``A fast iterative shrinkage-thresholding
  algorithm with application to wavelet-based image deblurring,'' in
  \emph{Proc. IEEE Int. Conf. Acoust., Speech and Signal Process. (ICASSP)},
  2009, pp. 693--696.

\bibitem{baraniuk_model_cs}
R.~G. Baraniuk, V.~Cevher, M.~F. Duarte, and C.~Hegde, ``Model-based
  compressive sensing,'' \emph{IEEE Trans. Info. Theory}, vol.~56, no.~4, pp.
  1982--2001, 2010.

\bibitem{distributed_deep_cs}
H.~Palangi, R.~Ward, and L.~Deng, ``Distributed compressive sensing: {A} deep
  learning approach,'' \emph{IEEE Trans. Signal Process.}, vol.~64, no.~17, pp.
  4504--4518, 2016.

\bibitem{lohit_deep_cs}
S.~Lohit, K.~Kulkarni, R.~Kerviche, P.~Turaga, and A.~Ashok, ``Convolutional
  neural networks for noniterative reconstruction of compressively sensed
  images,'' \emph{IEEE Trans. Comput. Imag.}, vol.~4, no.~3, pp. 326--340,
  2018.

\bibitem{lista}
K.~Gregor and Y.~LeCun, ``Learning fast approximations of sparse coding,'' in
  \emph{proc. Int. Conf. Machine Learning (ICML)}.\hskip 1em plus 0.5em minus
  0.4em\relax Madison, WI, USA: Omnipress, 2010, p. 399–406.

\bibitem{unrolling_mag}
V.~Monga, Y.~Li, and Y.~C. Eldar, ``Algorithm unrolling: Interpretable,
  efficient deep learning for signal and image processing,'' 2019.

\bibitem{amp_cs}
M.~Borgerding, P.~Schniter, and S.~Rangan, ``{AMP}-{I}nspired deep networks for
  sparse linear inverse problems,'' \emph{IEEE Trans. Signal Process.},
  vol.~65, no.~16, pp. 4293--4308, 2017.

\bibitem{pokala_firmnet}
P.~K. Pokala, A.~G. Mahurkar, and C.~S. Seelamantula, ``{FirmNet}: {A} sparsity
  amplified deep network for solving linear inverse problems,'' in \emph{Proc.
  IEEE Int. Conf. Acoust., Speech and Signal Process. (ICASSP)}, 2019, pp.
  2982--2986.

\bibitem{trainable_ista}
D.~Ito, S.~Takabe, and T.~Wadayama, ``Trainable {ISTA} for sparse signal
  recovery,'' \emph{IEEE Trans. Signal Process.}, vol.~67, no.~12, pp.
  3113--3125, 2019.

\bibitem{leung_icassp20}
V.~C.~H. Leung, J.-J. Huang, and P.~L. Dragotti, ``Reconstruction of {FRI}
  signals using deep neural network approaches,'' in \emph{Proc. IEEE Int.
  Conf. Acoust., Speech and Signal Process. (ICASSP)}, 2020, pp. 5430--5434.

\bibitem{leung_eusipco21}
V.~C.~H. Leung, J.-J. Huang, Y.~C. Eldar, and P.~L. Dragotti, ``Reconstruction
  of {FRI} signals using autoencoders with fixed decoders,'' in \emph{Proc.
  European Signal Process. Conf. (EUSIPCO)}, 2021.

\bibitem{Adcock2015}
B.~Adcock, A.~C. Hansen, and B.~Roman, ``The quest for optimal sampling:
  Computationally efficient, structure-exploiting measurements for compressed
  sensing,'' \emph{in Compressed Sensing and its Applications. Springer}, pp.
  143--167, 2015.

\bibitem{lustig_mri}
M.~Lustig, D.~Donoho, and J.~M. Pauly, ``Sparse {MRI}: The application of
  compressed sensing for rapid {MR} imaging,'' \emph{Magn. Reson. Med.},
  vol.~58, no.~6, p. 1182–1195, 2007.

\bibitem{ravishanker}
S.~Ravishankar and Y.~Bresler, ``Adaptive sampling design for compressed
  sensing {MRI},'' in \emph{Int. Conf. IEEE Eng. Medicine Biol. Soc.}, 2011,
  pp. 3751--3755.

\bibitem{oedipus}
J.~P. Haldar and D.~Kim, ``{OEDIPUS}: An experiment design framework for
  sparsity-constrained {MRI},'' \emph{IEEE Trans. Medical Imaging}, vol.~38,
  no.~7, pp. 1545--1558, 2019.

\bibitem{mulleti_radar}
S.~Mulleti, C.~Saha, H.~S. Dhillon, and Y.~C. Eldar, in \emph{Proc. IEEE Radar
  Conf. (RadarConf)}.

\bibitem{baldassarre}
L.~Baldassarre, Y.-H. Li, J.~Scarlett, B.~Gözcü, I.~Bogunovic, and V.~Cevher,
  ``Learning-based compressive subsampling,'' \emph{IEEE J. Selected Topics in
  Signal Process.}, vol.~10, no.~4, pp. 809--822, 2016.

\bibitem{gozcu}
B.~G\"ozc\"u, R.~K. {Mahabadi}, Y.~{Li}, E.~{Il\i cak}, T.~{\c{C}ukur},
  J.~{Scarlett}, and V.~{Cevher}, ``Learning-based compressive {MRI},''
  \emph{IEEE Trans. Medical Imag.}, vol.~37, no.~6, pp. 1394--1406, 2018.

\bibitem{candes_spmag}
E.~J. Candes and M.~B. Wakin, ``An introduction to compressive sampling,''
  \emph{IEEE Signal Process. Mag.}, vol.~25, no.~2, pp. 21--30, Mar. 2008.

\bibitem{candes_uncertainity}
E.~J. Candes, J.~Romberg, and T.~Tao, ``Robust uncertainty principles: {E}xact
  signal reconstruction from highly incomplete frequency information,''
  \emph{IEEE Trans. Info. Theory}, vol.~52, no.~2, pp. 489--509, Feb. 2006.

\bibitem{yi_mri}
\BIBentryALTinterwordspacing
K.~H. Jin, M.~Unser, and K.~M. Yi, ``Self-supervised deep active accelerated
  {MRI},'' \emph{CoRR}, vol. abs/1901.04547, 2019. [Online]. Available:
  \url{http://arxiv.org/abs/1901.04547}
\BIBentrySTDinterwordspacing

\bibitem{jmodl}
H.~K. Aggarwal and M.~Jacob, ``{J-MoDL}: {J}oint model-based deep learning for
  optimized sampling and reconstruction,'' \emph{IEEE J. Selected Topics in
  Signal Process.}, vol.~14, no.~6, pp. 1151--1162, 2020.

\bibitem{huijben2019learning}
I.~A.~M. Huijben, B.~S. Veeling, K.~Janse, M.~Mischi, and R.~J.~G. van Sloun,
  ``Learning sub-sampling and signal recovery with applications in ultrasound
  imaging,'' \emph{IEEE Trans Medical Imag.}, vol.~39, no.~12, pp. 3955--3966,
  2020.

\bibitem{prony}
G.~R. DeProny, ``Essai experimental et analytique: {S}ur les lois de la
  dilatabilit{\'e} de fluides {\'e}lastiques et sur celles de la force
  expansive de la vapeur de l'eau et de la vapeur de l'alcool, {\`a}
  diff{\'e}rentes temp{\'e}ratures,'' \emph{J. de l'Ecole polytechnique},
  vol.~1, no.~2, pp. 24--76, 1795.

\bibitem{sarkar_mp}
Y.~Hua and T.~K. Sarkar, ``Matrix pencil method for estimating parameters of
  exponentially damped/undamped sinusoids in noise,'' \emph{IEEE Trans.
  Acoust., Speech and Signal Process.}, vol.~38, no.~5, pp. 814--824, May 1990.

\bibitem{antennaSelectionViaCO}
S.~Joshi and S.~Boyd, ``Sensor selection via convex optimization,'' \emph{IEEE
  Trans. Signal Process.}, vol.~57, no.~2, pp. 451--462, 2009.

\bibitem{chepuri2015}
S.~P. {Chepuri} and G.~{Leus}, ``Sparsity-promoting sensor selection for
  non-linear measurement models,'' \emph{IEEE Trans. Signal Process.}, vol.~63,
  no.~3, pp. 684--698, 2015.

\bibitem{nemhauser}
G.~L. Nemhauser, L.~A. Wolsey, and M.~L. Fisher, ``An analysis of
  approximations for maximizing submodular set functions—i,'' \emph{Math.
  Program.}, vol.~14, no.~1, pp. 265--294, 1978.

\bibitem{antennaSelectionKnapsack}
H.~Godrich, A.~P. Petropulu, and H.~V. Poor, ``Sensor selection in distributed
  multiple-radar architectures for localization: {A} knapsack problem
  formulation,'' \emph{IEEE Trans. Signal Process.}, vol.~60, no.~1, pp.
  247--260, 2012.

\bibitem{fujishige2005submodular}
S.~Fujishige, \emph{Submodular Functions and Optimization}, ser. ISSN.\hskip
  1em plus 0.5em minus 0.4em\relax Elsevier Science, 2005.

\bibitem{mulleti_fdoct}
C.~S. Seelamantula and S.~Mulleti, ``Super-resolution reconstruction in
  frequency-domain optical-coherence tomography using the
  finite-rate-of-innovation principle,'' \emph{IEEE Trans. Signal Process.},
  vol.~62, no.~19, pp. 5020--5029, Oct. 2014.

\end{thebibliography}
	
\end{document}